\documentclass[12pt,a4paper]{article}
\usepackage{cite}
\usepackage{latexsym}
\usepackage{amsbsy}
\usepackage{amssymb}
\usepackage{epsfig}
\usepackage{rotate}
\newcounter{figures}
\newcounter{tables}

\newlength{\templength}
\newenvironment{equationarray}
{ \setlength{\templength}{\arraycolsep} \setlength{\arraycolsep}{2pt} 
\begin{eqnarray} }
{ \setlength{\arraycolsep}{\templength} \end{eqnarray} }
\newenvironment{equationarrayzero}
{ \setlength{\templength}{\arraycolsep} \setlength{\arraycolsep}{0pt} 
\begin{eqnarray} }
{ \setlength{\arraycolsep}{\templength} \end{eqnarray} }

\newcommand{\lsim}{\mathrel{\mathop{\kern 0pt \rlap
  {\raise.2ex\hbox{$<$}}}
  \lower.9ex\hbox{\kern-.190em $\sim$}}}
\newcommand{\gsim}{\mathrel{\mathop{\kern 0pt \rlap
  {\raise.2ex\hbox{$>$}}}
  \lower.9ex\hbox{\kern-.190em $\sim$}}}   
\begin{document}

\null


\vspace{0.5cm}


\begin{center}
\Large \bfseries
Neutrino masses, mixing and oscillations \footnote{
Lectures given at the {\em 1999 European School of High Energy Physics},
Casta Papiernicka, Slovakia, August 22--September 4, 1999}
\\[0.5cm]
\large \mdseries \upshape
S.M. Bilenky
\\[0.25cm]
\normalsize \itshape
Joint Institute for Nuclear Research, Dubna, Russia, \\
and
\\
Dipartimento di Fisica,
Universita' di Roma "Tor Vergata" \\
Via della Ricerca Scientifica 1 -- 00133 Roma, Italy 
\\[0.5cm]
\vspace{0.5cm}
\end{center}

\tableofcontents
\newpage

\section{Introduction}
\label{Introduction}
\setcounter{equation}{0}
\setcounter{figures}{0}
\setcounter{tables}{0}

During many years neutrino physics was a very important 
branch of elementary particle physics. In the last few years the 
interest to neutrinos particularly increased.
This is connected first of all with the success of
the Super-Kamiokande experiment in which very convincing 
evidence in favour of oscillations of atmospheric neutrinos were obtained.

It is plausible that tiny neutrino masses and neutrino
mixing are connected with the new large scale in physics.
This scale determines the smallness of neutrino masses
with respect to the masses of charged leptons and quarks.
In such a scenario neutrinos with definite masses are truly neutral 
Majorana 
particles (quarks and leptons have charges and are Dirac particles)
It is evident, however, that many new 
experiments are necessary to reveal the real origin of neutrino masses and 
mixing. 

Experimental neutrino physics is a very difficult and exciting 
field of research. Now it is a time when many new ideas and methods 
are being proposed. In CERN and other laboratories projects of 
new neutrino experiments are developing.
Possibilities of new neutrino facility, neutrino factory, are
investigated in different laboratories.
Thus it is a very appropriate time to
discuss neutrino physics at the CERN-JINR school.

In these lectures I will consider different possibilities of 
neutrino mixing. Then, I will discuss in some details neutrino 
oscillations in vacuum and in matter. In the last part of the 
lectures I will consider the present experimental situation. 

I tried to give in these lectures some important results
and details of derivation
of some results. I hope that lectures will be useful for those who want 
to study physics of massive neutrinos. More results 
and details can be found in the books
\cite{Mohapatra-Pal91}--\cite{Kim93} and reviews 
\cite{Bilenky-Pontecorvo78}--\cite{BGG}.

Most references to original papers can be found in\cite{BGG}

\section{Neutrino mixing}

According to the Standard Model of electroweak interaction the Lagrangian
of the interaction of neutrinos with other particles is given by the
Charged Current (CC) and the Neutral Current (NC) Lagrangians: 

\begin{equationarrayzero}
&&
\mathcal{L}_{I}^{\mathrm{CC}}
= - \frac{g}{2\sqrt{2}} \,
j^{\mathrm{CC}}_{\alpha} \, W^{\alpha}
+ \mathrm{h.c.} \,,
\label{CC}\\
&&
\mathcal{L}_{I}^{\mathrm{NC}}
= - \frac{g}{2\cos\theta_{W}} \,
j^{\mathrm{NC}}_{\alpha} \, Z^{\alpha} \,.
\label{NC}
\end{equationarrayzero}
Here $g$ is the electroweak interaction constant, $\theta_W$ is the weak 
(Weinberg) angle and $W^{\alpha}$ and $Z^{\alpha}$ are the fields of the
$W^{+-}$ and $Z^{0}$ vector bosons. If neutrino masses are equal to zero
in this case CC and NC interactions conserve electron
$L_e$, muon $L_{\mu}$ and tauon $L_{\tau}$ lepton numbers
 \begin{equationarrayzero}
&& \sum L_e =const, \sum L_\mu =const, \sum L_\tau =const
\label{cos}
\end{equationarrayzero}%
The values of the lepton numbers of charged leptons, neutrinos
and other particles are given in the Table \ref{lepton numbers}.

\begin{table}[t]
\begin{center}
Table \ref{lepton numbers}.
Lepton numbers of neutrinos and charged leptons.\\
Lepton numbers of all other particles are equal to zero.\\
\vspace{12pt}
\renewcommand{\arraystretch}{1.45}
\setlength{\tabcolsep}{0.5cm}
\begin{tabular}{|cccc|}
\hline
& $L_{e}$ & $L_{\mu}$ & $L_{\tau}$ \\
\hline
$\left( \nu_{e} \, , \, e^{-} \right)$
& $+1$ & 0 & 0 \\
$\left( \nu_{\mu} \, , \, \mu^{-} \right)$ &
0 & $+1$ & 0 \\
$\left( \nu_{\tau} \, , \, \tau^{-} \right)$ &
0 & 0 & $+1$ \\
\hline
\end{tabular}
\end{center}
\refstepcounter{tables}
\label{lepton numbers}
\end{table}

According to the \emph{neutrino mixing hypothesis} masses of neutrinos are
different from zero and \emph{neutrino mass term} does not conserve lepton
numbers. For the fields of $\nu_{lL}$ that enter into CC and NC
Lagrangians (\ref{CC}) and (\ref{NC})  we have, in this case,
\begin{equation}
\nu_{{l}L} = \sum_{i} U_{{l}i} \, \nu_{iL} \,
\label{mixing}
\end{equation}
where $\nu_i$ is the field of neutrino with mass $m_i$
and $U$ is the unitary mixing matrix.

The relation (\ref{mixing}) leads to violation of lepton numbers due to 
small neutrino mass differences and neutrino mixing. To reveal such effects 
special experiments (neutrino oscillation experiments, neutrinoless
double $\beta$-decay experiments and others) are necessary. We will
discuss such experiments later. Now we will consider {\it different
possibilities of neutrino mixing}.

Let us notice first of all that the relation \ref{mixing}
is similar to the analogous relation in the quark case. The standard
CC current of quarks have the form
\begin{equation}
j_\alpha^{CC} = 2 (\overline{u_L} \gamma_\alpha d'_L
+ \overline{c}_L \gamma_\alpha s'_L
+\overline{t}_L \gamma_\alpha b'_L)
\end{equation}
Here
\begin{equation}
d'_L = \sum_{q=d,s,b} V_{uq} q_L, 
\qquad s'_L = \sum_{q=d,s,b} V_{cq} q_L,
\qquad b'_L = \sum_{q=d,s,b} V_{tq} q_L
\end{equation}
where $V$ is Cabibbo--Kobayashi--Maskawa quark mixing matrix.
There can be, however, a fundamental difference between mixing of 
quarks and neutrino mixing. Quarks are charged four--component Dirac 
particles: quarks and antiquarks  have different charges.

For neutrinos with definite masses
there are two possibilities:
\begin{enumerate}
\item
 In case the total lepton number $L = L_e + L_\mu + L_\tau$
is conserved, neutrino with definite masses $\nu_i$
are four--component {\it Dirac particles} (neutrinos
and antineutrinos differ by the sign of $L$);
\item
If there are no conserved lepton numbers, neutrinos with definite 
masses $\nu_i$ are two-component {\it Majorana particles} 
(there are no quantum numbers in this case that can allow to 
distinguish neutrino from antineutrino).
\end{enumerate}

The nature of neutrino masses and the character of neutrino mixing is 
determined by the {\it neutrino mass term}.

\subsection{Dirac Neutrinos}

If the neutrino mass term is generated by the same standard 
Higgs mechanism, that is responsible for the mass generation
of quarks and charged leptons, then for the neutrino mass term we have
\begin{equation}
\mathcal{L}^{\mathrm{D}} = - \sum_{l,l'}
\overline{\nu_{{l'}R}} \, M^{\mathrm{D}}_{l'l} \, \nu_{{l}L} +
\mathrm{h.c.} \,
\label{Dirac}
\end{equation}
where $M^D$ is the complex $3 \times 3$ matrix and $\nu_{lR}$
is the right--handed singlet. In the case of mass term (\ref{Dirac})
 the total Lagrangian is invariant under global gauge invariance
\begin{equation}
\nu_{lL} \to e^{i \alpha} \nu_{lL}\,,
\qquad \nu_{lR} \to e^{i \alpha} \nu_{lR} \,,
\qquad {l} \to e^{i \alpha} {l} \,,
\label{global gauge transformations}
\end{equation}
where $\alpha$ is a constant that does not depend on the flavor index 
$l$. The invariance under the transformation (\ref{global gauge 
transformations})  means that  the total lepton number 
$L = L_e + L_\mu + L_\tau$  is conserved
\begin{equation}
\sum L = const
\label{Diracsum}
\end{equation}
Now let us diagonalize the mass term (\ref{Dirac}). The complex matrix 
$M^\mathrm{D}$ can be diagonalized by biunitary transformation
\begin{equation}
M^{\mathrm{D}} = V m U^\dagger \,,
\label{complex matrix}
\end{equation}
where $ V^{\dagger} V = 1$, $U^{\dagger} U =1$ and
$m_{ik} = m_i \delta_{ik} ,\quad m_i > 0$.

With the help of (\ref{complex matrix}), from (\ref{Dirac}) for the neutrino 
mass term we obtain the standard expression
\begin{equation}
\mathcal{L}^{\mathrm{D}} = - \sum_{l',l,i} \overline{\nu_{{l'}R}}\,
V^{}_{l'i} \, m_{i}(U^{\dagger})_{il} \,\nu_{{l}L} +
\mathrm{h.c.} =
- \sum_{{i=1}}^3 m_i \overline{\nu_{i}}\nu_{i}
\label{Dirac1}
\end{equation}
Here 
$$\nu_i = \nu_{iL} + \nu_{iR} \hskip1cm          (i = 1,2,3)$$
and
$$\nu_{iL} = \Sigma_l (U^{\dagger})_{il} \nu_{lL}$$
$$\nu_{iR} = \Sigma_l (V^{\dagger})_{il} \nu_{lR}$$  
For the neutrino mixing we have
\begin{equation}
\nu_{{l}L} = \sum_{i} U_{{l}i} \, \nu_{iL} \,
\label{mixing1}
\end{equation}

Processes in which the total lepton number is conserved, 
like $\mu \rightarrow e + \gamma $ and others, are,
 in principle, allowed in the case of mixing of Dirac massive neutrinos.
 It can be shown, however, that the probabilities of such processes 
are much smaller than the experimental upper bounds. 

Neutrinoless double $\beta$--decay,
$$ (A,Z) \rightarrow (A,Z+2) + e^- +e^-\, ,$$
due to the conservation of the total lepton number is forbidden
in the case of Dirac massive neutrinos.

\subsection{Majorana neutrinos}

Neutrino mass terms that are generated in the framework of the models 
beyond the Standard Model, like the Grand Unified SO(10) Model, do not
conserve lepton numbers $L_e$, $L_\mu$ and $L_\tau$.
Let us build the most general neutrino mass term that does not conserve
$L_e$, $L_\mu$ and $L_\tau$. 

Neutrino mass term is a linear combination of the products of
left--handed and right-handed components of neutrino fields.
Notice that $(\nu_L)^C =C (\overline\nu_L)^T$ is the right--handed
component and $(\nu_R)^C =C (\overline\nu_R)^T$ is the
left--handed component.
Here C is  the charge conjugation matrix, that satisfies
the relations $C \gamma_{\alpha}^T C^{-1} = - \gamma_{\alpha}$,
$C^T = - C$,\,  $C^{\dagger} C = 1$
\footnote{In fact, $L$ and $R$ components satisfy the relations 
\[ \frac{1+\gamma_5}{2} \nu_{L} = 0
\qquad \frac{1-\gamma_5}{2}  \nu_{R} = 0 \]
From the first of these relations we have
$\overline{\nu}_L (1-\gamma_5)/2 =0$.
Further, from this last relation we obtain 
$[(1-\gamma_5)/2]^T \overline{\nu}_L^T = 0.$
Multiplying this relation by the matrix $C$ from the left  and
taking into account that $C\gamma_5^TC^{-1} = \gamma^5$
we have $[(1-\gamma_5)/2] (\nu_L)^C = 0$. Thus, $(\nu_L)^C$ is
right--handed component. Analogously we can show that  $(\nu_R)^C$ 
is left-handed component.}

The most general Lorentz--invariant neutrino mass term 
in which flavor neutrino fields $\nu_{lL}$ and 
right--handed singlet fields $\nu_{lR}$ enter has 
the following form
\begin{equation}
\mathcal{L}^{\mathrm{D-M}} = - \frac{1}{2}\overline{(n_L)^C}\,
M  n_L + \mathrm{h.c.}
\label{LDM2}
\end{equation}
Here 
\begin{equation}
n_L = \left(\begin{array}{c} \displaystyle \nu_L'
\\ \displaystyle (\nu_R')^C \end{array} \right)
\quad \mbox{with} \quad 
\nu_L' = \left( \begin{array}{c} \displaystyle \nu_{eL}
\\ \displaystyle \nu_{{\mu}L} \\
\displaystyle \nu_{{\tau}L} \end{array} \right)
\quad \mbox{and} \quad
\nu_R' = \left( \begin{array}{c} \displaystyle \nu_{eR}
\\ \displaystyle \nu_{{\mu}R} \\
\displaystyle \nu_{{\tau}R} \end{array} \right) \,,
\label{nL}
\end{equation}
$M$ is complex $6\times 6$ matrix. Taking into account that 
$\overline{(\nu_L)^C} = - \nu_L^T C^{-1}$ we have 
\begin{equation}
\mathcal{L}^{\mathrm{D-M}} = \frac{1}{2} n_L^T \,
\mathcal{C}^{-1}\, M \, n_L + \mathrm{h.c.} \,.
\label{LDM22}
\end{equation}
From this expression it is obvious that there is no
global gauge invariance in the case of the mass term (\ref{LDM2}), 
i.e. that the mass term (\ref{LDM2}) does not conserve lepton numbers.

The matrix $M$ is symmetric. In fact, taking into account the 
commutation properties of fermion fields we have 
\begin{equation}
n_L^T \,\mathcal{C}^{-1}\, M \, n_L = - n_L^T \,
(\mathcal{C}^{T})^{-1} \, M^T \, n_L =
n_L^T \, \mathcal{C}^{-1} \, M^T \, n_L \,.
\end{equation}
From this relation it follows that
 $$M^T =M$$
The symmetric $6\times 6$ matrix can be presented in the form
\begin{equation}
M = \left( \begin{array}{cc} \displaystyle M_{L}
\null & \null \displaystyle (M_{\mathrm{D}})^T \\
\displaystyle M_{\mathrm{D}} \null & \null \displaystyle M_{R}
\end{array} \right) \,.
\label{two02}
\end{equation}
where  $M_L = M_L^T$, $M_R = M_R^T$ and $M^D$ are $3\times 3$ matrices. 
With the help of (\ref{two02}) for the mass term (\ref{LDM22}) we have
\begin{equation}
\mathcal{L}^{\mathrm{D-M}} = 
\mathcal{L}^{\mathrm{M}}_L + \mathcal{L}^{\mathrm{D}} +
\mathcal{L}^{\mathrm{M}}_R \,.
\label{LDM21}
\end{equation}
Here ${\cal L}^D$ is the Dirac mass term, that we have considered before,
and the new terms
\begin{equation}
\mathcal{L}^{\mathrm{M}}_L = - \frac{1}{2} \sum_{l',l}
\overline{(\nu_{{l'}L})^c} \, M^{L}_{l'l} \, \nu_{{l}L}
+ \mathrm{h.c.} \,,
\end{equation}
\begin{equation}
\mathcal{L}^{\mathrm{M}}_R = - \frac{1}{2} \sum_{l',l}
\overline{(\nu_{{l'}R})^c} \, M^{R}_{l'l} \, \nu_{{l}R} 
+ \mathrm{h.c.} \,,
\end{equation}
which do not conserve lepton numbers are called left--handed and 
right--handed Majorana mass terms, respectively.
The mass term (\ref{LDM2}) is called Dirac--Majorana mass term.

A symmetrical matrix can be diagonalized with the help of unitary 
transformation
$$M = (U^{\dagger})^T m U^{\dagger} \, .$$
Here $U$ is unitary matrix and $m_{ik} = m_i \delta_{ik},\, m_i>0$.
Using the relation (11) we can write the mass term (\ref{LDM22})
in the standard form 
\begin{equation}
\mathcal{L}^{\mathrm{D-M}} = - \frac {1}{2}
(\overline{U^{\dagger} n_L})^C \, m U^+ n_L + \mathrm{h.c.} 
= - \frac {1}{2} \overline{\nu} m \nu = - \frac {1}{2}
\sum_{i=1}^6 m_i \overline{\nu}_i \nu_i \,,
\label{Majorana mass1}
\end{equation}
where 
\begin{equation}
\nu = U^+ n_L+ (U^+ n_L)^C = \left( \begin{array}{c} \displaystyle
\nu_1 \\ \displaystyle  \nu_2 \\ \displaystyle \vdots \\ \displaystyle 
\nu_6 \end{array} \right) \,,
\label{Majorana mass2}
\end{equation}
Thus the fields $\nu_i$ (i=1,2...6) are the fields of neutrinos with 
mass $m_i$.
From (\ref{Majorana mass2}) it follows that the fields $\nu_i$ satisfy
the {\it Majorana condition}
\begin{equation}
\nu_i^C = \nu_i \,,
\label{Majorana mass3}
\end{equation}

Let us obtain now the relation that connects the left--handed flavor
fields $\nu_{lL}$ with the massive fields $\nu_{iL}$.
From (\ref{Majorana mass2}) for the left--handed components  we have
\begin{equation}
n_L = U \nu_L. 
\end{equation}
From this relation for the flavor field $\nu_{lL}$ it follows
\begin{equation}
\nu_{lL} = \sum_{i=1}^6 U_{li} \nu_{iL} \qquad
(l=e,\mu,\tau) \,,
\label{Majorana mass22}
\end{equation}
Thus, in the case of Dirac--Majorana mass term, the flavor fields 
are linear combinations of left--handed components of
six massive Majorana fields. From (\ref{Majorana mass22}) it follows
that the fields $\nu_{lR}^C$ are orthogonal
linear combinations of the same massive Majorana fields
\begin{equation}
(\nu_{lR})^C = \sum_{i=1}^6 U_{\overline{l}i} \nu_{iL} \,.
\end{equation}
In the case of Majorana field  particles and antiparticles, 
quanta of the field, are identical.
 In fact, for fermion fields $\nu(x)$ we have in general case
\begin{equation}
\nu(x) = \int {\frac {1} {(2\pi)^{3/2}} \, \frac {1} {\sqrt{2p^0}} \,
\left[c_r(p) u^r(p)e^{-ipx} + d_r^{\dagger}(p) C
\left(\overline{u}^r(p)\right)^{T} e^{ipx}\right] d^3 p}
\end{equation}
where $c_r(p) (d_r^{\dagger}(p))$ is the operator of absorption  
of particle (creation of antiparticle) with momentum  $p$ and
helicity $r$. If the field $\nu(x)$ satisfies the  Majorana condition
(\ref{Majorana mass3}), then we have
\begin{equation}
c_r(p) = d_r(p)
\end{equation}
Let us stress that it is natural that the neutrinos with definite masses
in the case of Dirac--Majorana mass term are Majorana neutrinos: in fact
there are no  conserved quantum numbers that could allow us
to distinguish particles and antiparticles.

\subsection{The simplest case of one generation (Majorana neutrinos)}

It is instructive to consider in detail the Dirac--Majorana mass term in 
the simplest case of one generation. We have
\begin{equationarray}
\mathcal{L}^{\mathrm{D-M}} \null & \null = \null & \null
- \frac{1}{2} \, m_{L} \, (\overline{\nu}_L)^c \, \nu_L -
m_{\mathrm{D}} \, \overline{\nu}_R \, \nu_L - \frac{1}{2} \, m_{R}\,
\overline{\nu}_R \, (\nu_R)^c + \mathrm{h.c.}
\nonumber \\
\null & \null = \null & \null - \frac{1}{2} \,
(\overline{n}_L)^c \, M \, n_L + \mathrm{h.c.}\,,
\label{two01}
\end{equationarray}
where
\begin{equation}
n_L \equiv \left(\begin{array}{c} \displaystyle \nu_L \\
\displaystyle (\nu_R)^c \end{array} \right) \,,
\qquad M \equiv \left( \begin{array}{cc} \displaystyle m_{L}
\null & \null \displaystyle m_{\mathrm{D}} \\ \displaystyle
m_{\mathrm{D}} \null & \null \displaystyle m_{R} \end{array} \right) \,.
\end{equation}

Let us assume that the parameters $m_L, m_R$ and $m_D$ are real
(the case of CP invariance). 
In order to diagonalize the mass term (\ref{two01})
let us write the matrix $M$ in the form
\begin{equation}
M = \frac{1}{2} \, \mathrm{Tr}M + \underline{M} \,,
\label{two03}
\end{equation}
where Tr~$M = m_L + m_{D}$ and
\begin{equation}
\underline{M}= \left( \begin{array}{cc} \displaystyle
- \frac{1}{2} ( m_R - m_L ) \null & \null \displaystyle
m_{\mathrm{D}} \\ \displaystyle m_{\mathrm{D}}
\null & \null \displaystyle \frac{1}{2} ( m_R - m_L ) \end{array}\right)\,.
\label{two04}
\end{equation}
For the symmetrical real matrix we have
\begin{equation}
\underline{M} = \mathcal{O} \, \underline{m}^{} \mathcal{O}^T \,.
\label{two06}
\end{equation}
Here
\begin{equation}
\mathcal{O} = \left(\begin{array}{rr} \displaystyle
\cos\vartheta \null & \null \displaystyle \sin\vartheta
\\ \displaystyle - \sin\vartheta \null & \null \displaystyle
\cos\vartheta \end{array}\right)
\label{two07}
\end{equation}
is an orthogonal matrix, and 
$\underline{m}_{ik} = \underline{m}_{i} \delta_{ik}$, 
where
\begin{equation}
\underline{m}_{1,2} =  \mp \frac{1} {2}
\sqrt{(m_R - m_L)^2 + 4 m_{D}^2}
\label{recov01}
\end{equation}
are eigenvalues of the matrix $\underline{M}$.

From (\ref{two06}),  (\ref{two07}) and (\ref{recov01}) for the parameters 
$\cos\vartheta$ and $\sin\vartheta$ we easily find the following
expressions
\begin{equation}
\cos 2\vartheta = 
\frac{m_R-m_L}{\sqrt{ ( m_R - m_L )^2 + 4 \, m_{D}^2 }} \,, \quad
\tan 2\vartheta = \frac{2m_D}{ ( m_R - m_L )} \,.
\label{two08}
\end{equation}
For the matrix $M$ from (\ref{two06}) and  (\ref{recov01}) we have
$$M = O m' O^T$$
where
\begin{equation}
m'_{1,2} = \frac{1}{2} \,
( m_R + m_L ) \mp \sqrt{( m_R  - m_L )^2 + 4 \, m_{D}^2 } \,.
\label{two05}
\end{equation}
The eigenvalues $m_i'$ can be positive or negative. Let us 
write 
\begin{equation}
m'_i  =  m_i \eta_i \,,
\label{two061}
\end{equation}
where $m_i = |m_i|$ and $\eta_i$ is the sign of the i-eigenvalue.
With the help of (\ref{two06}) and (\ref{two061}) we have
$$M = (U^{\dagger})^T m U^{\dagger}$$
Here
$$U^{\dagger} = \sqrt{\eta} O^T$$
where $\sqrt{\eta}$ takes  the values 1 and $i$.

Now using the general formulas (\ref{Majorana mass1}) and 
(\ref{Majorana mass2}) for the mass term we have
\begin{equation}
{\cal L}^{D-M} = - \frac {1} {2} \sum_{i=1,2} m_i 
\overline{\nu}_i \nu_i
\end{equation}
Here $\nu_i = \nu_i^C$ is the field of the Majorana particles with mass 
$m_i$. The fields $\nu_L$ and $(\nu_R)^C$ are connected with 
massive fields by the relation
\begin{equation}
\left(\begin{array}{c} \displaystyle \nu_L \\ \displaystyle
(\nu_R)^C \end{array} \right)  = U \left(
\begin{array}{c} \displaystyle \nu_{1L} \\ \displaystyle
\nu_{2L} \end{array}\right)\,,
\label{nLw}
\end{equation}
where $U = O(\sqrt{\eta})^*$ is a $2\times 2$ mixing matrix.

Let us consider now three special cases. 
\begin{enumerate}
\item No mixing \\
Assume $m_D$ = 0. In this case $\theta=0$, $m_1 = m_L$, $m_2 = m_R$
and $\eta = 1$ (assuming that $m_L$ and $m_R$ are positive). 
From (\ref{nLw}) we have
\begin{equation}
\nu_L = \nu_{1L} \qquad (\nu_R)^C = \nu_{2L} \,.
\end{equation}
Thus, if $m_{\mathrm{D}} = 0$ there is no mixing. 
For the Majorana fields $\nu_1$ and $\nu_2$ we have
\begin{equation}
\nu_1 = \nu_{L} +(\nu_L)^C \,
\end{equation}
\begin{equation}
\nu_2 = \nu_{R} +(\nu_R)^C \,.
\end{equation}
\item  Maximal mixing \\
Assume $m_R = m_L$,  $m_{D} \neq 0$.  From Eq. (\ref{two08}),
(\ref{two05}) and (\ref{nLw}) we have
\begin{equation}
\theta = \frac {\pi} {4}, \qquad
m_{1,2} = m_L \mp m_{D} \,.
\end{equation}
(assuming $|m_{D}| < m_L$) and
\begin{equation}
\nu_L = \frac  {1}{\sqrt{2}} \nu_{1L} +   \frac  {1}{\sqrt{2}} \nu_{2L};
\qquad
(\nu_R)^C = - \frac  {1}{\sqrt{2}} \nu_{1L} +
\frac  {1}{\sqrt{2}} \nu_{2L} \,.
\end{equation}
Thus if the diagonal elements of the mass matrix $M$ are equal, then
 we have maximal mixing.

\item See--saw mechanism of neutrino mass generation \\
Assume $m_L = 0$ and
\begin{equation}
m_D \ll m_R
\label{rr01}
\end{equation}
From (\ref{recov01}) and (\ref{two05}) we have in this case
\begin{equation}
m_1 \simeq \frac {m_D^2} {m_R}, 
\quad m_2 \simeq m_R,
\quad \theta \simeq  \frac {m_D} {m_R}
\quad (\eta_1=-1, \eta_2=1)\,.
\end{equation}

Neglecting terms linear in $ {m_D}/ {m_R} \ll 1 $,
from (\ref{nLw}) we have 
\begin{equation}
\nu_L \simeq -i \nu_{1L},
\qquad (\nu_R)^C  \simeq \nu_{2L} \,.
\end{equation}
For the Majorana fields we have
\begin{equation}
\nu_1 \simeq i  \nu_{L} - i(\nu_{L})^C,
\qquad \nu_2 = \nu_{R} + (\nu_{R})^C \,.
\end{equation}
Thus if the condition (\ref{rr01}) is satisfied, in the spectrum 
of masses of Majorana particles there are one light particle with the mass 
$m_1 << m_D$ and one heavy particle with the mass
$m_1 >> m_D$. The condition $m_L = 0$ means that the lepton 
number is violated only by the right--handed term $- \frac {1} {2} m_R 
\overline{\nu}_R (\nu_R)^C$ that is characterized 
by the large mass $m_R$. It is natural to assume that the parameter
$m_{D}$ which characterizes the Dirac term $-m_{D} \overline{\nu}_R \nu_L$
is of the order of lepton or quark masses. The mass of the light Majorana 
neutrino $m_1$ will be in this case much smaller than the mass of
lepton or quark. This is famous {\it see-saw mechanism}.
This mechanism connects the smallness of the neutrino masses
with respect to the masses
of other fundamental fermions with violation of the lepton numbers
at very large scale (usually $m_{\it D} \simeq M_{GUT} \simeq 10^{16}$~GeV. 

In the case of the see--saw for three families in the spectrum of
masses of Majorana particles there are three light masses 
$m_1, m_2, m_3$ (masses of neutrinos)
and three very heavy masses $M_1, M_2, M_3$. 
Masses of neutrinos are connected with the masses of heavy Majorana
particles by the see--saw relation
\begin{equation}
m_i  \simeq \frac {(m^i_{\mathrm{f}})^2} {M_i} \ll  m^i_{\mathrm{f}}
\qquad (i=1,2,3) \,.
\end{equation}
where $m_{\mathrm{f}}^i$ is the mass of lepton or quark in $i$-family.
The see--saw mechanism is a plausible explanation of the experimentally 
observed smallness of neutrino masses.
Let us stress that if neutrino masses are of the see-saw origin then \\
\noindent a. neutrinos with definite masses are Majorana particles; \\
\noindent b. there are three massive neutrinos; \\
\noindent c. there must be a hierarchy of neutrino masses
$m_1 \ll m_2 \ll m_3$.
\end{enumerate}

\section{Neutrino oscillations}

The most important consequences of the neutrino mixing are so called 
{\it neutrino oscillations}.
Neutrino oscillations were first considered  by B.~Pontecorvo 
 many years ago in 1957-58. Only one type of neutrino was known  
at that time and there was general belief that neutrino is a massless
two--component particle. B.~Pontecorvo draw attention that there is no 
known principle which requires neutrino to be massless (like gauge 
invariance for the photon)
and that the investigation of neutrino oscillations is a very sensitive 
method to search for effects of small neutrino masses. 
We will consider here in detail the phenomenon of neutrino oscillations.

Assume that there is neutrino mixing
\begin{equation}
\nu_{{\alpha}L} = \sum_{i} U_{{\alpha}i} \, \nu_{iL} \,.
\label{mix01}
\end{equation}
where $U^{\dagger} U = 1$ and $\nu_i$ is the field of neutrino 
(Dirac or Majorana) with the mass $m_i$.
The field $\nu_{\alpha L} $ in (\ref{mix01}) are flavor fields 
($\alpha = e, \mu, \tau$) and in general also sterile ones
($\alpha = s_1,...$).
Let us assume that neutrino mass differences are small and 
different neutrino masses cannot be resolved in neutrino 
production and detection processes. 

For the state of neutrino with momentum $\vec{p}$  we have
\begin{equation}
|\nu_\alpha\rangle = \sum_i U_{{\alpha}i}^* \, |\nu_i\rangle \,.
\label{state}
\end{equation}
where $|\nu_i\rangle$ is the vector of state of neutrino with
momentum $\vec{p}$, energy
\begin{equation}
E_i = \sqrt{p^2 + m_i^2 } \simeq p + \frac{ m_i^2 }{ 2 p }
\qquad (p\gg m_i) \,.
\label{energy}
\end{equation}
and (up to the terms  ${m_i^2}/ {p^2}$) helicity is equal to -1. If
at the initial time $t=0$ the state of neutrino is  $|\nu_\alpha\rangle$
at the time $t$ for the neutrino state we have 
\begin{equation}
|\nu_\alpha\rangle_t =
\sum_i U_{{\alpha}i}^* \, e^{ - i E_i t } \, |\nu_i\rangle \,.
\label{state-t}
\end{equation}
The vector  $|\nu_\alpha\rangle$ is the superposition of the states of 
all types of neutrino.
In fact, from (\ref{state}), using unitarity of the mixing matrix, we have
\begin{equation}
|\nu_i\rangle = \sum_{\alpha'} |\nu_{\alpha'}\rangle U_{{\alpha'}i}\,.
\label{state-t1}
\end{equation}
From (\ref{state-t}) and (\ref{state-t1}) we have 
\begin{equation}
|\nu_\alpha\rangle_t =
\sum_{\alpha'} |\nu_{\alpha'}\rangle
{\cal A}_{\nu_\alpha';{\nu_\alpha}}(t) \,.
\label{state-t2}
\end{equation}
where
\begin{equation}
 {\cal A}_{\nu_\alpha'; \nu_\alpha}(t) = \sum_i
 U_{\alpha' i}e^{-iE_it}U_{\alpha i}^* \,.
\label{state-t30}
\end{equation}
is the amplitude of the transition $\nu_\alpha \rightarrow \nu_{\alpha'}$
at the time $t$.  The transition amplitude
${\cal A}_{\alpha';\alpha}(t)$ has a simple meaning: the term 
$U^*_{\alpha i}$ is the amplitude of the transition from the state
$|\nu_\alpha\rangle$ to the state  $|\nu_i\rangle$;
the term  $e^{-iE_it}$ describes the evolution in the state with energy
$E_i$; the term $U_{\alpha' i}$ is the transition amplitude from the state
$|\nu_i\rangle$ to the state   $|\nu_\alpha'\rangle$.

The different  $|\nu_i\rangle$ gives \underline{coherent} contribution to 
the amplitude ${\cal A}_{\nu_\alpha'; \nu_\alpha}(t)$. 
From (\ref{state-t30}) it follows that the  transitions between 
different states can take place only if: i) at least two neutrino masses 
are different; ii) the mixing matrix is non--diagonal.
In fact, if all neutrino masses are equal  we have  
$a(t) = e^{-iEt} \sum U_{\alpha' i} U_{\alpha i}^* = e^{-iEt} 
\delta_{\alpha'\alpha}$.
If the mixing matrix is diagonal (no mixing), we have  
${\cal A}_{\nu_\alpha'; \nu_\alpha}(t) = e^{-i E_\alpha t } 
\delta_{\alpha'\alpha}$.

Let us numerate neutrino masses in such a way that $m_1<m_2<...<m_n$.
For the transition 
probability, from (\ref{state-t30}), we have the following expression:
\begin{equationarray}
P_{\nu_\alpha\to\nu_{\alpha'}} &&= \left|
\sum_{i} U_{{\alpha'}i} \, \left[\left(e^{ - i(E_i -E_1)t} - 1\right)
 +1\right] \, U_{{\alpha}i}^* \right|^2 
\label{prob2}\\
&&\quad = \left| \delta_{\alpha{\alpha'}} +
 \sum_{i} U_{{\alpha'} i} \, U_{{\alpha}i}^*
 \left( e^{- i \,
 \Delta{m}^2_{i1}\, \frac {L}{2 p}} - 1 \right)\right|^2 \,,
\nonumber
\end{equationarray}
where $\Delta m^2_{i1} = m_i^2 - m^2_1$ and $L \simeq t$
is the distance between neutrino source and neutrino detector.
Thus the neutrino transition probability 
depends on the ratio $\frac {L} {E}$, 
the range of values of which is determined by the conditions 
of an experiment.

It follows from Eq. (\ref{prob2}) that the transition probability depends 
in the general case on $(n-1)$ neutrino mass squared differences and
parameters that characterize the mixing matrix $U$. The $n\times n$
matrix $U$ is characterized by  $n_\theta = {n(n-1)}/ {2}$
angles. The number of phases for Dirac and Majorana cases is different.
If neutrino with definite masses $\nu_i$ are Dirac particles the 
number of phases is equal to  $n_\phi^D ={(n-1)(n-2)}/ {2}$. 
If $\nu_i$ are Majorana particles the number of phases is equal to
 $n_\phi^{M_j} =  {n(n-1)}/ {2}$.

Notice that from (\ref{prob2}) it follows that transition probability is 
invariant under the transformation
\begin{equation}
U_{{\alpha}i} \to e^{-i \beta_\alpha} \, U_{{\alpha}i} \, e^{i\alpha_i}
\label{phasetrafo}
\end{equation}
where $\beta_{\alpha}$ and $\alpha_i$ are arbitrary real phases. 
From (\ref{phasetrafo}) it follows that the number of 
phases that enter into the transition probability is equal to
$n_\phi =  {(n-1)(n-2)}/ {2}$ in both Dirac and Majorana cases. 
We come to the 
conclusion that additional Majorana phases do not enter into the
transition probability . Thus, by investigation of neutrino oscillations
it is impossible to distinguish the case of Dirac neutrinos from 
the case of Majorana neutrinos.

Let us consider now oscillations of antineutrinos. For the 
vector of state of antineutrino with momentum  $\vec{p}$
from (\ref{mix01}) we have
\begin{equation}
|\overline{\nu}_\alpha\rangle =
\sum_{i} U_{{\alpha}i}|\overline{\nu}_{i}\rangle \qquad
({\mathrm{Dirac\,\, case}})
\label{state-t3}
\end{equation}
\begin{equation}
|\overline{\nu}_\alpha\rangle =
\sum_{i} U_{{\alpha}i}|\nu_{i}\rangle \qquad
({\mathrm{Majorana\,\, case}})
\label{state-t4}
\end{equation}
where  $|\overline{\nu}_i\rangle$ ($|\nu_i\rangle$)
is the state of antineutrino (neutrino) with momentum  
$\vec{p}$, energy  $E_i = \sqrt{p^2 +m_i^2} \simeq p + {m_i^2}/{2p}$
and helicity equal to +1 ( up to ${m_i^2}/ {p^2}$ terms).

In analogy with (\ref{state-t30}) for the amplitude of the transition
$\overline{\nu}_\alpha \to \overline{\nu}_{\alpha'}$
in both Dirac and Majorana cases we have
\begin{equation}
{\cal A}_{\overline{\nu}_{\alpha'}; \overline{\nu}_{\alpha}}(t) = 
\sum_i U_{\alpha'i}^* e^{-iE_it}U_{{\alpha}i} \,.
\label{state-tt}
\end{equation}
If we compare (\ref{state-t30}) and (\ref{state-tt}) we come to the
conclusion that 
\begin{equation}
{\cal A}_{\overline{\nu}_{\alpha'}; \overline{\nu}_{\alpha}}(t) =
{\cal A}_{\nu_\alpha; \nu_{\alpha'}}(t) \,.
\label{state-tt1}
\end{equation}
Thus for the transition probabilities we have the following relation 
 \begin{equation}
{\mathrm P}(\nu_\alpha \to \nu_{\alpha'}) =
{\mathrm P}(\overline{\nu}_{\alpha'} \to \overline{\nu}_\alpha) \,.
\label{state-tz}
\end{equation}

This relation is the consequence of CPT invariance. 
If CP invariance in the lepton sector takes place then
for Dirac neutrinos we have 
\begin{equation}
U_{{\alpha}i}^*  = U_{{\alpha}i}
\label{state-tw}
\end{equation}
while for Majorana neutrinos, from CP invariance, we have
\begin{equation}
U_{{\alpha}i}\eta_i  = U_{{\alpha}i}^*\, ;
\label{state-tzM}
\end{equation}
where $\eta_i = \pm i$ is the CP parity of the Majorana neutrino with 
mass $m_i$. From (\ref{state-t30}), (\ref{state-tt1}), (\ref{state-tw})
and (\ref{state-tzM}) it follows that 
in case of CP invariance we have 
\begin{equation}
{\mathrm P}(\nu_\alpha \to \nu_\alpha') =
{\mathrm P}(\overline{\nu}_{\alpha} \to \overline{\nu}_\alpha') \,.
\label{state-tzz}
\end{equation}

Let us go back to the Eq. (\ref{prob2}). It is 
obvious from (\ref{prob2}) that if the conditions of 
an experiment are such that $\Delta m^2_{i 1} \frac {L} {p} \ll 1$
for all $i$ then neutrino oscillations cannot be observed. To observe
neutrino oscillations it is necessary that for at least one neutrino 
mass squared difference the condition  $\Delta m^2 \frac {L} {p} \gsim 1$ 
is satisfied. We will discuss this condition later. 

\subsection{Two neutrino oscillations}

Let us consider in details the simplest case of the oscillations between
two neutrinos  $\nu_\alpha \leftrightarrows \nu_{\alpha'}$
($\alpha' \neq \alpha; \alpha,\alpha'$  are equal to $\mu$, e or $\tau$,
$\mu$,...).
The index $i$ in Eq. (\ref{prob2}) takes values 1 and 2 and for the 
transition probability we have
\begin{equation}
{\mathrm P}(\nu_\alpha \to \nu_\alpha') =
|\delta_{{\alpha'}\alpha} + U_{\alpha'_2}  U_{\alpha_2}^*
 (e^{- i \Delta m^2_{2 1} \frac {L} {2p}} -1)|^2 
\label{state-tzzw}
\end{equation}
For  $\alpha' \neq \alpha$  we have from (\ref{state-tzzw})
\begin{equation}
{\mathrm P}(\nu_\alpha \to \nu_\alpha') = 
{\mathrm P}(\nu_{\alpha'} \to \nu_\alpha)
= \frac {1} {2} {\mathrm A}_{{\alpha'}\alpha} (1 - \cos \Delta m^2 
\frac {L} {2p})
\label{state-tzzz}
\end{equation}
Here the amplitude of oscillations is equal to
\begin{equation}
{\mathrm A}_{\alpha'; \alpha} = 4 |U_{{\alpha'}2}|^2  
|U_{{\alpha}2}|^2
\end{equation}
and $\Delta m^2 = m^2_2 - m_1^2$.
 Due to unitarity of the mixing matrix 
\begin{equation}
|U_{{\alpha}2}|^2 + |U_{{\alpha'}2}|^2 = 1
\qquad (\alpha' \neq \alpha)
\end{equation}

Let us introduce the mixing angle $\theta$ 
\begin{equation}
|U_{{\alpha}2}|^2 = \sin^2 \theta
\qquad |U_{{\alpha'}2}|^2 = \cos^2 \theta
\end{equation}
Thus the oscillation amplitude ${\mathrm A}_{{\alpha'}; \alpha}$
 is equal to 
\begin{equation}
{\mathrm A}_{\alpha'; \alpha} = \sin^2 2\theta 
\end{equation}
The survival probabilities ${\mathrm P}(\nu_\alpha \to \nu_\alpha)$
and  ${\mathrm P}(\nu_{\alpha'} \to \nu_{\alpha'})$  can be 
obtained from (\ref{state-tzzw})  or from the condition of the conservation 
of the total probability   ${\mathrm P}(\nu_\alpha \to \nu_\alpha) +
{\mathrm P}(\nu_{\alpha} \to \nu_{\alpha'}) = 1$. 
We have
\begin{equation}
{\mathrm P}(\nu_\alpha \to \nu_\alpha) = {\mathrm P}(\nu_{\alpha'} \to 
\nu_{\alpha'}) =
 1 - \frac {1}{2} \sin^2 2\theta (1 - \cos \frac {\Delta m^2 L} {2p})
\label{wanda}
\end{equation}
Thus in the case of two neutrinos the transition 
probabilities are characterized by two parameters  $\sin^2 2 \theta$ and 
$\Delta m^2$.

Let us notice that in the case of transitions between two neutrinos 
only moduli of the elements of the mixing matrix enter into expressions 
for the transition probabilities. This means that in this case
the CP relation (\ref{state-tz}) is satisfied automatically. Thus, in order 
to observe effects of CP violation in the lepton sector
the transitions between three neutrinos must take place (this is 
similar to the quark case: for two families of quarks CP is conserved 
due to unitarity of the mixing matrix).

We also notice that the expression (\ref{state-tzzz}) for the transition
probability can be written in the form
\begin{equation}
{\mathrm P}(\nu_\alpha \to \nu_{\alpha'}) = \frac {1}{2} \sin^2 2\theta 
\left(1 - \cos 2\pi \frac {L}{L_0}\right)
\end{equation}
where 
\begin{equation}
L_0 =  4 \pi \frac {E} {\Delta m}
\end{equation}
is the oscillation  length.
The expression (\ref{state-tzzz}) is written in the units $\hbar = c = 1$.
We can write it in the form 
\begin{equation}
{\mathrm P}(\nu_\alpha \to \nu_{\alpha'}) = \frac {1}{2} \sin^2 2\theta
\left(1 - \cos\,  2.54 \Delta m^2\frac {E} {L}\right)
\end{equation}
where $\Delta m^2$ is neutrino mass squared difference
in eV$^2$, $L$ is the distance in m (km) and $E$ is the neutrino energy in
MeV (GeV).
For the oscillation length we have
\begin{equation}
L_0 =  2.47  \frac {E ({\mathrm{MeV}})}{\Delta m^2 ({\mathrm{eV}}^2)}
\,{\mathrm{ m}}
\end{equation}

The Eq. (\ref{state-tzzz}) and (\ref{wanda}) describe periodical 
transitions (oscillations) between different types of neutrinos due to 
difference of neutrino masses and  to neutrino mixing.
The transition probability depends periodically on $ {L}/{E}$.
At the values of $ {L}/ {E}$ at which the condition
$2.54\,  {\Delta m^2}({L}/{E}) = \pi (2n + 1)\quad (n=0,1,...)$
is satisfied, the transition probability is equal to 
the maximal value $\sin^2 2\theta $.
If the condition $2.54 \, {\Delta m^2}({L}/ {E}) = 2 \pi n$
is satisfied, the transition probability is equal to zero.

In order to see neutrino oscillations it is necessary that the
parameter $\Delta m^2$ is large enough so that the condition 
$\Delta m^2 ({L}/ {E}) \ge 1$ 
is satisfied. This condition allows us to estimate the minimal 
value of the parameter $\Delta m^2$ that can be revealed in an 
experiment on the search for neutrino oscillations.
For short and long baseline experiments with accelerator
(reactor) neutrinos for $\Delta m^2_{min}$ we have, respectively 
10 -- 1 eV$^2$, $10^{-2}$ -- 10$^{-3}$ eV$^2$ (10$^{-1}$
 -- 10$^{-2}$ eV$^2$, 10$^{-2}$ -- 10$^{-3}$ eV$^2)$. 
For atmospheric and solar neutrinos for 
$\Delta m^2_{min}$ we have 10$^{-2}$ -- 10$^{-3}$ eV$^2$ and
10$^{-10}$ -- 10$^{-11}$ eV$^2 $ , respectively. 
Let us notice that in the case of $\Delta m^2 ({L}/{E})\ll 1$,
due to averaging over neutrino spectrum and over distances 
between neutrino production and detection points,
 the term $\cos \Delta m^2 ({L}/{2p})$
in the transition probability disappears and the averaged 
transition probabilities are given by 
$\overline{{\mathrm P}}(\nu_\alpha \to \nu_{\alpha'}) = \frac{1}{2}
\sin^2 2\theta$ and 
$\overline{{\mathrm P}}(\nu_\alpha \to \nu_{\alpha}) = 1-\frac{1}{2} 
\sin^2 2\theta$.

\subsection{Three neutrino oscillations in the case of neutrino\\
 mass hierarchy}

The two neutrino transition probabilities (\ref{state-tzzz}) 
and (\ref{wanda}) are usually used for the analysis of experimental 
data. Let us consider now the case of the transitions 
between three flavor neutrinos. 

General expressions for transition probabilities between three
neutrino types are characterized by 6 parameters and have a  rather 
complicated form.
We will consider the case of hierarchy of neutrino masses 
$$ m_1 \ll m_2 \ll m_3 $$
which corresponds to the oscillations of solar and 
atmospheric neutrinos (we have in mind that $\Delta m^2_{21}$
can be relevant for oscillations of solar neutrinos and $\Delta m^2_{31}$
can be relevant for oscillations of atmospheric neutrinos;
from the analysis of the experimental data it follows that
$\Delta m^2_{sol} \simeq 10^{-5}$~eV$^2$ (or $10^{-10}$ eV$^2$)
and $\Delta m^2_{atm} \simeq 10^{-3}$~eV$^2$; see later).
We will see that transition probabilities have in this case the
rather simple two--neutrino form.

Let us consider neutrino oscillations in experiments for
which the largest neutrino mass squared difference $\Delta m^2_{31}$
is relevant. For such experiments 
\begin{equation}
\Delta m^2_{1 2} \frac {L} {2p}\ll 1
\end{equation}
and for the probability of the transition $\nu_\alpha \to \nu_{\alpha'}$,
from (\ref{prob2}) we obtain the following expression
\begin{equation}
{\mathrm P}(\nu_\alpha \to \nu_{\alpha'}) 
= \left|\delta_{{\alpha'}\alpha} + U_{{\alpha'}3}U_{{\alpha}3}^*
\left(e^{{- i \Delta m^2_{3 1}\, \frac{L}{2p}}}-1\right)\right|^2
\label{state-tttz}
\end{equation}
For the transition probability $\nu_\alpha \to \nu_{\alpha'}$
 $(\alpha' \neq \alpha)$ from (\ref{state-tttz}) we have 
\begin{equation}
{\mathrm P}(\nu_\alpha \to \nu_\alpha') 
=  \frac {1} {2} {\mathrm A}_{\alpha';\alpha} 
\left(1 - \cos \Delta m^2_{3 1}\, \frac {L} {2p}\right)
\label{state-tttw}
\end{equation}
where the amplititude of oscillations is given by 
\begin{equation}
{\mathrm A}_{\alpha';\alpha} = 4 |U_{{\alpha'}3}|^2 |U_{{\alpha}3}|^2
\label{state-ttww}
\end{equation}

Using unitarity of the mixing matrix, for the survival probability we 
obtain, from (\ref{state-tttw}) and (\ref{state-ttww}),
\begin{equation}
{\mathrm P}(\nu_\alpha \to \nu_\alpha) 
=  1 - \sum_{{\alpha}'\ne\alpha} {\mathrm P}(\nu_\alpha \to \nu_{\alpha'})
= 1 - \frac {1}{2} {\mathrm B}_{\alpha;\alpha}
\left(1 - \cos \Delta m^2_{3 1} \frac{L}{2p}\right)
\label{state-tvw}
\end{equation}
where 
\begin{equation}
{\mathrm B}_{\alpha;\alpha} = 4 |U_{{\alpha}3}|^2 (1 -|U_{{\alpha}3}|^2)
\end{equation}
It is natural that  Eq. (\ref{state-tttw}) and (\ref{state-ttww})
 have the same dependence on 
the parameter ${L}/{E}$ as the standard two--neutrino formulas 
(\ref{state-tzzw}) and (\ref{wanda}): only  the largest $\Delta m^2$ is
relevant for the oscillations.
The oscillation amplitudes  ${\mathrm A}_{\alpha;\alpha}$ and 
${\mathrm B}_{\alpha;\alpha}$  depend on the 
moduli squared  of the mixing matrix elements that connect 
neutrino flavors with the heaviest neutrino $\nu_3$.
Further, from the unitarity of the mixing matrix it follows that 
\begin{equation}
 |U_{{e}3}|^2 +  |U_{{\mu}3}|^2 + |U_{{\tau}3}|^2 = 1
\end{equation}
Thus, in three--neutrino case with hierarchy of neutrino masses, the
transition probabilities in experiments for which $\Delta m_{31}^2$
is relevant are described by three parameters: 
$\Delta m^2_{31}$, $|U_{e3}|^2$ and $|U_{\mu3}|^2$ (remember that in 
the two neutrino case there are two parameters, $\Delta m^2$
and $\sin^22\theta$ ).

Since only moduli of the elements of the mixing matrix enter into 
transition probabilities, the relation 
\begin{equation}
{\mathrm P}(\nu_\alpha \to \nu_{\alpha'}) = {\mathrm P}
(\overline{\nu}_\alpha \to \overline{\nu}_{\alpha'}) 
\end{equation}
holds (as in the two--neutrino case). Thus the violation of the 
CP--invariance in the lepton sector
cannot be revealed in the case of three neutrinos with mass hierarchy.
Notice that the relation 
\begin{equation}
{\mathrm P}(\nu_\alpha \to \nu_{\alpha}) 
= {\mathrm P}(\nu_{\alpha'} \to \nu_{\alpha'}) \, ,
\end{equation}
which takes place in the case of two neutrino 
oscillations, is not valid in the three--neutrino case.

Let us consider now neutrino oscillations in the case of experiments
for which $\Delta m^2_{21}$ is relevant 
($\Delta m^2_{2 1}\, \frac {L} {2p} \gsim 1$).
From (\ref{state-t30}) for the survival probability we obtain in this case 
the following expression
\begin{equation}
{\mathrm P}(\nu_\alpha \to \nu_{\alpha}) = 
\left|\sum_{i=1,2}|U_{{\alpha}i}|^2 
e^{ - i \Delta m^2_{{i}1} \frac {L}{2p}}
+ |U_{{\alpha}3}|^2 
e^{ - i \Delta m^2_{{3}1} \frac {L} {2p}}\right|^2
\label{zita}
\end{equation}
Due to averaging over neutrino spectra and source--detector distances, the 
interference term $\cos \Delta m^2_{{3}1}\,( {L}/ {2p})$
in Eq. (\ref{zita}) disappears and for the probability we have
\begin{equation}
{\mathrm P}(\nu_\alpha \to \nu_{\alpha}) = 
|\sum_{i=1,2}|U_{{\alpha}i}|^2 e^{- i \Delta m^2_{{i}1}
\frac {L} {2p}}|^2 + |U_{{\alpha}3}|^4
\label{zitta}
\end{equation}

Further, from the unitarity relation $\sum_{i=1}^3 |U_{{\alpha}i}|^2 = 1$
we have
\begin{equation}
\sum_{i=1,2}|U_{{\alpha}i}|^4  = 
(1 - |U_{{\alpha}3}|^2)^2 - 2 |U_{{\alpha}1}|^2  |U_{{\alpha}2}|^2 
\label{zotta}
\end{equation}
Using (\ref{zotta}) we can present the survival probability in the form 
\begin{equation}
{\mathrm P}(\nu_\alpha \to \nu_{\alpha}) = 
(1 - |U_{{\alpha}3}|^2)^2{\mathrm P}^{(1,2)}(\nu_\alpha \to 
\nu_{\alpha}) +  |U_{{\alpha}3}|^4
\label{zista}
\end{equation}
Here
\begin{equation}
{\mathrm P}^{(1,2)}(\nu_\alpha \to \nu_{\alpha}) = 1 - \frac{1}{2} 
\sin^2 2 \overline{\theta}_{1 2} 
(1 - \cos^2 \Delta m^2_{2 1}\, \frac{L}{2p})
\label{zista1}
\end{equation}
and the angle $\overline{\theta}_{1 2}$  is determined by the relations
\begin{equation}
\cos^2 \overline{\theta}_{1 2} = \frac {|U_{{\alpha}1}|^2} 
{\sum_{i=1,2} |U_{{\alpha}i}|^2},
\qquad \sin^2 \overline{\theta}_{1 2} = 
\frac {|U_{{\alpha}2}|^2} {\sum_{i=1,2} |U_{{\alpha}i}|^2},
\end{equation}
The probability ${\mathrm P}^{(1,2)}(\nu_e \to \nu_e)$
 has the two--neutrino form and 
it is characterized by two parameters: $\Delta m^2_{31}$ and
$\sin^2 2\overline{\theta}_{1 2}$.
We have derived the expression (\ref{zista1}) for the case of 
the oscillations in 
vacuum. Let us notice that similar expression is valid for the case of 
the neutrino transitions in matter.

The expressions (\ref{state-tttw}), (\ref{state-tvw}) and (\ref{zista1})
can be used to describe neutrino oscillations in atmospheric
and long baseline neutrino experiments (LBL) as well as 
in solar neutrino experiments. 
In the framework of neutrino mass hierarchy, in the probabilities of 
transition of atmospheric (LBL) and solar neutrinos enter different 
$\Delta m^2$ ($\Delta m^2_{31}$ and $\Delta m^2_{2,1}$, respectively)
and the only element that connects oscillations of atmospheric (LBL)
and solar neutrinos is $|U_{e3}|^2$.
From LBL reactor experiment CHOOZ and Super--Kamiokande experiment
it follows that this element is small (see later). This means that 
oscillations of atmospheric (LBL) and solar neutrinos  
are described by different elements of the neutrino mixing matrix.

\section{Neutrino in matter}

Up to now we have considered oscillations of neutrinos in vacuum. 
If there is neutrino mixing the effects of the matter can significantly
enhance the probability of the transitions between different types of 
neutrinos (MSW effect). We will consider here this effect in some
details. 

Let consider neutrinos with momentum $\vec{p}$.
The equation of the motion for a free neutrino has the form
\begin{equation}\
i \frac{\partial |\psi(t)\rangle} {\partial t} = H_0 |\psi(t)\rangle
\label{mot}
\end{equation}
Let us develop the state $|\psi(t)\rangle$ over states of neutrinos 
with definite flavor $|\nu_\alpha\rangle$ ($\alpha = e, \mu, \tau$).
We have
\begin{equation}\label{vac}
|\psi(t)\rangle = \sum_\alpha |\nu_\alpha\rangle{a}_\alpha(t)
\end{equation}
where $a_\alpha(t)$ is the wave function of neutrino in the 
flavor representation. From (\ref{mot}) for $a_\alpha(t)$ we obtain 
the equation
\begin{equation}
i \frac{\partial a_\alpha(t)} {\partial t} = 
\sum_{\alpha'} \langle\nu_\alpha | H_0 | \nu_{\alpha'}\rangle
a_{\alpha'}(t)
\label{mot1}
\end{equation}

Now we will develop the state $|\nu_\alpha\rangle$ over the eigenstates 
$|\nu_i\rangle$  of the free Hamiltonian H$_0$:
\begin{equationarrayzero}
&& H_0 |\nu_i\rangle = E_i |\nu_i\rangle\,,\\
&& E_i = \sqrt{p^2 + m_i^2} \simeq p + \frac {m_i^2} {2p}\,.
\end{equationarrayzero}%
We have:
\begin{equation}\label{motts}
|\nu_\alpha\rangle = \sum_i |\nu_i\rangle \langle\nu_i|\nu_\alpha\rangle
\end{equation}
If we compare (\ref{motts}) and (\ref{state}) we find 
\begin{equation}
\langle\nu_i|\nu_\alpha\rangle = U_{{\alpha}i}^*
\qquad
\langle\nu_\alpha|\nu_i\rangle = U_{{\alpha}i}
\end{equation}

Further we have
\begin{equation}\label{motz}
 \langle\nu_\alpha|H_0|\nu_{\alpha'}\rangle = 
\sum_i \langle\nu_\alpha|\nu_i\rangle 
\langle\nu_i|H_0|\nu_i\rangle 
\langle\nu_i|\nu_{\alpha'}\rangle
= \sum_i U_{{\alpha}i} \frac {m^2_i} {2p} U^{\dagger}_{i{\alpha'}} + 
p \delta_{\alpha{\alpha'}}
\end{equation}
The last term of \ref{motz}, which is proportional to unit matrix, cannot 
change the flavor state of neutrino.
This term  can be excluded from the equation of  motion by  
redefining  the phase of the function $a(t)$. We have:
\begin{equation}\label{motww}
 i \frac {\partial a(t)} {\partial t} = U \frac {m^2} {2p} U^\dagger  a(t)
\end{equation}
This equation can be easily solved. 
Let us multiply (\ref{motww}) by the matrix $U^{\dagger}$ from the left.
Taking into account unitarity of the mixing matrix we have:
\begin{equation}\label{motwv}
 i \frac {\partial a'(t)} {\partial t} =  \frac {m^2} {2p} a'(t)
\end{equation}
where $a'(t) = U^{\dagger} a(t)$.
The solution of equation (\ref{motwv}) has the form
\begin{equation}\label{motvv}
a'(t) = e^{- i \frac {\Delta m^2} {2p} t} a'(0)\, .
\end{equation}
For the function $a(t)$ in flavor representation, 
from (\ref{motwv}) and (\ref{motvv}), we find
\begin{equation}
\label{motss}
a(t) = U e^{- i \frac {\Delta m^2} {2p} t}U^{\dagger} a(0)
\end{equation}
and  for the amplitude of the $\nu_\alpha \to \nu_{\alpha'}$
transition in vacuum from (\ref{motss}) we obtain the expression 
\begin{equation}\label{motrr}
{\cal{A}}_{\nu_{\alpha'};\nu_\alpha}(t) = \sum_i U_{{\alpha'}i}
 e^{- i \frac {\Delta m^2_i} {2p}} U^*_{{\alpha}i}
\end{equation}
which (up to the irrelevant factor $e^{-ipt}$) coincides with 
(\ref{state-t30}).

Let us now introduce the effective Hamiltonian of interaction of 
flavor neutrino with matter. Due to coherent scattering of neutrino in 
matter, the refraction index of neutrino is given by the following
classical expression:
\begin{equation}\label{motrrs}
n(x)  = 1 + \frac {2\pi} {p^2} \mathrm{f}(0) \rho(x)
\end{equation}
Here $\mathrm{f}(0)$ is the amplitude of elastic neutrino
scattering in forward direction, and $\rho(x)$ is the 
number density of matter (the axis $x$ is the direction of $\vec{p}$). 
The effective interaction of neutrinos with matter is determined by the 
second term of Eq. (\ref{motrrs}) :
\begin{equation}\label{motri}
H_I(x) = p [n(x) - 1] = \frac{2\pi} {p} \mathrm{f}(0) \rho(x)
\end{equation}

NC scattering of neutrinos on electrons and nucleons 
(due to the Z-exchange) cannot change the flavor state of neutrinos.
This is connected with  $\nu_e - \nu_\mu - \nu_\tau$ universality of NC:
the corresponding effective Hamiltonian is proportional to the unit 
matrix\footnote{
Let us notice that if there are flavour and sterile neutrinos NC 
interactions with matter must be taken into account.}.

CC interaction (due to the W-exchange) gives contribution  only
 to the amplitude of the elastic $\nu_e$ -$e$ scattering 
\begin{equation}\label{acr}
\nu_e + e \to \nu_e + e
\end{equation}
For the corresponding effective Hamiltonian we have
\begin{equation}\label{mori}
{\cal H}_I(x) = \frac{G_F} {\sqrt{2}} 2 \overline{\nu}_{{e}L} \gamma^\alpha 
\nu_{{e}L} 
\overline{e} \gamma_\alpha (1 - \gamma_5) e  + h.c.
\end{equation}
The amplitude of process (\ref{acr}) is given by
\begin{equation}\label{mosi}
\mathrm{f}_{{\nu_e}e} = \frac {1} {\sqrt{2}\pi} G_F p
\end{equation}
and, from (\ref{motri}) and (\ref{mosi}), for the effective Hamiltonian 
in flavor representation we have
\begin{equation}\label{morsi}
H_I(x) = \sqrt{2} G_F \rho_e(x) \beta
\end{equation}
where $(\beta)_{\nu_e;\nu_e} = 1$, while all other elements of the matrix 
$\beta$ are equal to zero and $\rho_e(x)$ is the electron number density at 
the point $x$. 

The effective Hamiltonian of the neutrino interaction  with matter can
be also obtained by calculating of the average value of the Hamiltonian
(\ref{mori}) in the state which
 describes matter and neutrino with momentum $\vec{p}$
and negative helicity . Taking into account
that for non--polarized media 
\begin{equationarrayzero}
&&
\langle{\mathrm{mat}}\, | \overline{e}(\vec{x}) \gamma^\alpha e(\vec{x})
 |\,{\mathrm{ mat}}\rangle = \rho_e(\vec{x})\delta_{{\alpha}0}\,,\\
&&
\langle{\mathrm{mat}} | \overline{e}(\vec{x}) \gamma^\alpha \gamma_5 
e(\vec{x}) |{\mathrm{ mat}}\rangle = 0\,,
\end{equationarrayzero}%
from (\ref{mori}) we obtain (\ref{morsi}).

The evolution equation of neutrino in matter can be written, 
from (\ref{motww}) and (\ref{morsi}), in the following form ($t=x$):
\begin{equation}\label{morse}
i \frac {\partial a(x)} {\partial x} = (U \frac {m^2} {2p} U^{\dagger} 
+ \sqrt{2} G_F \rho_e(x) \beta) a(x)
\end{equation}

Let consider in detail the simplest case of two flavor neutrinos
(say, $\nu_e$ and $\nu_{\mu}$). In this case we have 
\begin{equation}
U = \left(
\begin{array}{rr} \displaystyle
\cos\vartheta \null & \null \displaystyle \sin\vartheta\\ 
\displaystyle - \sin\vartheta \null & \null 
\displaystyle \cos\vartheta \end{array}\right)
\label{two07w}
\end{equation}
where $\theta$ is the mixing angle. Further it is convenient 
to write the Hamiltonian in the form
\begin{equation}\label{mozze}
H = \frac {1} {2} {\mathrm{Tr}} H + H^m
\end{equation}
where Tr~$H = \frac {1} {2p} (m^2_1 + m^2_2) + \sqrt{2} G_F \rho_e$.
The first term of (\ref{mozze}), which is proportional to the unit matrix,
can be omitted. For the Hamiltonian  we have then
\begin{equation}
H^m(x)
= \frac {1} {4p}\left(\begin{array}{cc} \displaystyle
- \Delta m^2 \cos 2 \vartheta + A(x) &  \displaystyle
\Delta m^2 \sin 2 \vartheta\\ 
\displaystyle \Delta m^2 \sin 2 \vartheta & \displaystyle
\Delta m^2 \cos 2 \vartheta - A(x) \end{array}\right)
\label{two08w}
\end{equation}
where $\Delta m^2 = m^2_2 - m^2_1$ and 
$A(x) = 2 \sqrt{2} G_F \rho_e(x) p$.
The effect of matter is described by the quantity $A(x)$.
Notice that this quantity enters only into the diagonal elements of 
the Hamiltonian and has the dimensions of $M^2$.

Let us first consider the case of  constant density. 
In order to solve equation of motion we will diagonalize the 
Hamiltonian. We have:
\begin{equation}\label{morre}
H^m =  U^m E^m {U^{m}}^{\dagger}
\end{equation}
where $E^m_i$ is the eigenvalue of the matrix $H^m$ and
\begin{equation}
U^m = \left( \begin{array}{rr} \displaystyle
\cos \vartheta^m \null & \null \displaystyle
\sin\vartheta^m \\ 
\displaystyle - \sin\vartheta^m \null & 
\null \displaystyle \cos \vartheta^m \end{array}\right)
\label{two09}
\end{equation}
It is easy to see that 
\begin{equation}
E^m_{1,2} = \mp \frac {1} {4p} \sqrt{(\Delta m^2 \cos 2\theta - A)^2 + 
(\Delta m^2 \sin 2\theta)^2}.
\label{two11}
\end{equation}
Now, with the help of Eq. (\ref{morre}) -- (\ref{two11}), for the angle 
$\theta^m$ we have 
\begin{equation}
\tan 2\theta^m = \frac {\Delta m^2 \sin 2\theta} 
{\Delta m^2 \cos 2\theta - A};
\qquad
\cos 2\theta^m = \frac {\Delta m^2 \cos 2\theta - A} 
{\sqrt{(\Delta m^2 \cos 2\theta - A)^2 + (\Delta m^2 \sin 2\theta)^2}}
\label{two12}
\end{equation}
The states of flavor neutrinos are given by
\begin{equation}
|\nu_e\rangle = \cos\theta^m |\nu_{1 m}\rangle + \sin\theta^m 
|\nu_{2 m}\rangle; \qquad
|\nu_\mu\rangle = - \sin\theta^m |\nu_{1 m}\rangle 
+ \cos\theta^m |\nu_{2 m}\rangle
\label{two13}
\end{equation}
where $|\nu_{im}\rangle$ ($i=1,2$) are eigenvectors of the Hamiltonian of 
neutrino in matter and $\theta^m$  is the mixing angle of neutrino in 
matter. 

The solution of the evolution equation 
\begin{equation}
i \frac {\partial a(x)} {\partial x} = H_m a(x)
\label{two14}
\end{equation}
can be now easily found. With the help of (\ref{morre}) we have
\begin{equation}
i \frac {\partial a'(x)} {\partial x} = E^m a'(x)
\label{two15}
\end{equation}
where 
\begin{equation}
a'(x) = (U^m)^{\dagger} a(x).
\label{two51}
\end{equation}
The equation (\ref{two15}) has the following solution:
\begin{equation}
a'(x) = e^{- i E^m(x-x_0)}a'(x_0)
\label{two52}
\end{equation}
where $x_0$ is the point where the neutrino was produced.
Finally, from (\ref{two51}) and (\ref{two52}), we have 
\begin{equation}
a(x) = U^m e^{- i E^m(x-x_0)} (U^m)^{\dagger} a(x_0) 
\label{two21}
\end{equation}

The amplitude of the $\nu_\alpha \to \nu_{\alpha'}$ transition
in matter turns out to be
\begin{equation}
{\cal A}_{\nu_{\alpha'};\nu_\alpha} = \sum_{i=1,2} 
U^m_{{\alpha'}i} e^{- i E_i^m(x-x_0)}  U^*_{{\alpha}i}
\label{two22}
\end{equation}
and, from (\ref{two22}) and (\ref{two09}), we obtain the following  
transition probabilities, in full analogy with the two--neutrino 
vacuum case:
\begin{equationarrayzero}
&&
P^m(\nu_e \to \nu_\mu) =
P^m(\nu_\mu \to \nu_e) =
\frac {1} {2} \sin^2 2\theta ^m (1-\cos \Delta E^m L)\,,\\
&&
P^m(\nu_e \to \nu_e) =
P^m(\nu_\mu \to \nu_\mu) =
(1 - P^m (\nu_e \to \nu_\mu)
\label{two23}\,.
\end{equationarrayzero}
Here $\Delta E^m = E_2^m - E_1^m = \frac {1} {2p}
\sqrt{(\Delta m^2 \cos 2 \theta - A)^2 + 
(\Delta m^2 \sin 2 \theta )^2}$ and $L = x- x_0$ is the distance 
that neutrino passes in matter. 

For the oscillation length of neutrino in matter with constant density 
we have
\begin{equation}
L_0^m = 4 \pi \frac {p} {\sqrt{
(\Delta m^2 \cos 2\theta - A)^2 + (\Delta m^2 \sin 2\theta)^2}}
\end{equation}
The mixing angle and oscillation length in matter can differ
significantly from the vacuum values. It follows from (\ref{two12}) 
that if the condition\footnote{Eq. (\ref{two23}) is the condition at which 
the diagonal elements of the Hamiltonian of neutrino in matter vanish.
It is evident that in such a case the mixing is maximal.}
\begin{equation}
\Delta m^2 \cos 2\theta = A = 2 \sqrt{2} G_F \rho_e p
\label{two24}
\end{equation}
is satisfied, the mixing in matter is maximal 
($\theta^m ={\pi}/{4}$)
independently on the value of the vacuum mixing angle $\theta$.
Notice also that if the condition (\ref{two24}) is satisfied,
the distance between the energy levels of neutrinos in matter 
is minimal and the oscillation length in matter is maximal. We have
\begin{equation}
L_0^m = \frac {L_0} {\sin 2\theta}
\end{equation}
where $L_0 = 4 \pi {p}/({\Delta m})$ is the oscillation length in
vacuum. If the distance $L$ in the transition probabilities (\ref{two23}) 
is large (as in the Sun case) the effect of $\nu_e \to \nu_\mu$ 
 transitions is large even in case of a small 
vacuum mixing angle $\theta$. The relation (\ref{two24}) is called 
resonance condition.

The density of electrons in the Sun is not constant. It is maximal in 
the center of the Sun and decreases practically exponentially 
to its periphery.
The consideration of the dependence of $\rho_e$ on $x$ allowed to 
discover possibilities for the large effects of the transitions
of solar $\nu_e$'s into other states in matter (MSW effect).

Let us consider the evolution equation when
the Hamiltonian depends on the distance $x$
that neutrino passes in matter
\begin{equation}
i \frac {\partial a(x)} {\partial x} = {\it H}^{m}(x) a(x)
\label{two25}
\end{equation}
The Hermitian  Hamiltonian ${\it H}^{m}(x)$ can be diagonalized 
by a unitary transformation
\begin{equation}
{\it H}^{m}(x) = U^m(x) E^m(x) {U^{m}}^{\dagger}(x)
\label{two26}
\end{equation} 
where $U^m(x) {U^{m}}^{\dagger}(x) =1$ and $E_i^m$(x)  are eigenvalues 
of ${\it H}^{m}(x)$. From (\ref{two25}) and (\ref{two26}) we have 
\begin{equation}
{U^{m}}^{\dagger}(x) i \frac {\partial a(x)} {\partial t} = E^m(x) a'(x)
\end{equation}
where 
\begin{equation}
a'(x) ={ U^{m}}^{\dagger}(x) a(x)
\label{two27}
\end{equation}
Further, by taking into account that 
\begin{equation}
{U^{m}}^{\dagger}(x)  i  \frac {\partial a(x)} {\partial x} = 
i \frac {\partial a'(x)} {\partial x}+ i {U^{m}}^{\dagger}(x) 
\frac {\partial U^m(x)} {\partial x} a'(x)  \, ,
\end{equation}
we have the following equation for $a'(x)$:
\begin{equation}
i  \frac {\partial a'(x)} {\partial x} = 
\left( E^m(x) - i {U^{m}}^{\dagger}(x) 
\frac{\partial U^m(x)}{\partial x}\right)a'(x)\, .
\label{two29}
\end{equation}
In the case $\rho_e$ = const the equation (\ref{two29})
 coincides with (\ref{two15}).

Let us now assume that the function  $\rho_e(x)$ depends weakly on $x$
and the second term in Eq. (\ref{two27}) can be dropped (adiabatic 
approximation). It is evident that the solution of the equation 
\begin{equation}
i  \frac {\partial a'_i(x)} {\partial x} = 
E^m_i(x) a'_i(x)
\label{two30}
\end{equation}
has the form 
\begin{equation}
a'_i(x) = e^{\displaystyle - i \int_{x_0}^x{E_i^m(x)\, dx}} a'_i(x_0)
\label{two31}
\end{equation}
($x_0$ being the initial point).

It follows from (\ref{two30}) and (\ref{two31}) that, in the adiabatic 
approximation, a neutrino on the way from the point $x_0$ to the point
$x$ remains in the same energy level. 
From (\ref{two27}) and (\ref{two31}) we obtain the following solution 
of the evolution equation in flavor representation:
\begin{equation}
a(x) = U^m(x) e^{- i \int_{x_0}^x{E^m(x)\, dx}}
 {U^{m}}^{\dagger}(x_0)A(X_0)\,.
\end{equation}
Moreover the amplitude of $\nu_\alpha \to \nu_{\alpha'}$  transition 
in adiabatic approximation is given by
\begin{equation}\label{motrrm}
{\cal A}_{\nu_{\alpha'};\nu_\alpha} = \sum U^m_{{\alpha'}i}(x)
e^{- i \int_{x_0}^x{E_i^m(x)\, dx}} U^{m*}_{{\alpha}i}(x_0)\, .
\label{two32}
\end{equation}
The latter is similar to the expressions (\ref{motrr})
and (\ref{two22}) for the amplitudes of transition in vacuum and in
matter with $\rho_e$ = const.

For the case of the two flavor neutrinos
\begin{equation}
U^m(x) = \left(\begin{array}{rr} \displaystyle
\cos\vartheta^m(x)\null & \null \displaystyle
\sin\vartheta^m(x) \\
\displaystyle - \sin\vartheta^m(x) \null & \null \displaystyle
\cos\vartheta^m(x) \end{array}\right)
\label{two35}
\end{equation}
and $\tan 2\theta^m(x)$  and $cos 2\theta^m(x)$ are given by
Eq.~(\ref{two12}) in which
\begin{equation}\label{motrrw}
\mathrm{A}(x) = 2 \sqrt{2} G_F \rho_e(x) p
\end{equation}
The eigenvalues of the Hamiltonian $H^m(x)$ are given by 
Eq. (\ref{two11}). From (\ref{two35}) we have 
\begin{equation}
{U^{m}}^{\dagger}(x)\, \frac {\partial U^m(x)} {\partial x} =
\left( \begin{array}{cc} \displaystyle 0  & \displaystyle
\frac {\partial\theta^m(x)} {\partial x}\\
\displaystyle - \frac {\partial\theta^m(x)} {\partial x} &
\displaystyle 0 \end{array} \right)
\label{two36}
\end{equation}
and the exact equation (\ref{two29}) takes the form 
\begin{equation}
i\frac{\partial}{\partial x}
\left(\begin{array}{c} a'_1\\a'_2\end{array}\right)
=\left(\begin{array}{cc} E_1^m & \displaystyle
-i\frac{\partial\theta^m}{\partial x}\\
\displaystyle i\frac{\partial\theta^m}{\partial x}
& E_2^m \end{array}\right)
\left(\begin{array}{c} a'_1\\a'_2\end{array}\right)
\label{www1}
\end{equation}
The Hamiltonian ${ H^m}$ in the right--hand side 
of this equation can be written in the form
\begin{equation}
H_m = \frac{1}{2}(E^m_1+E^m_2) +
\left(\begin{array}{cc}-\frac{1}{2} \Delta E^m & 
\displaystyle -i\frac{\partial\theta^m}{\partial x}\\
\displaystyle i\frac{\partial\theta^m}{\partial x}
& \frac{1}{2}\Delta E^m \end{array}\right)
\label{www2}
\end{equation}
where $\Delta E^m = E^m_2 - E^m_1$. As we stressed several 
times, the term of the Hamiltonian which is proportional to the
unit matrix is not important for flavor evolution.

From Eq.~(\ref{www2}) it follows that adiabatic approximation is 
valid if the condition
\begin{equation}
\left| \frac {\partial \theta^m} {\partial x}\right| \ll
\frac {1} {2} \Delta E^m
\label{two37}
\end{equation}
is satisfied. With the help of (\ref{two12}) it is easy
to show that (\ref{two37}) can be written in the form
\begin{equation}
4 \sqrt{2} G_F p^2 \Delta m^2 \sin 2\theta 
\left|\frac {\partial \rho_e} {\partial x}\right| \ll
\left[(\Delta m^2 \cos 2\theta - A)^2 +
(\Delta m^2 \sin 2\theta)^2\right]^{3/2}.
\end{equation}
If the resonance condition
\begin{equation}
\Delta m^2 \cos 2\theta = A(x_R)
\label{two99}
\end{equation}
is satisfied at the point $x = x_R$, the condition of validity 
of the adiabatic approximation can be written in the form 
\begin{equation}
\frac {2 p \cos 2\theta \left|\frac {\partial}{\partial x} 
\ln \rho_e(x_R)\right|} {\Delta m^2 \sin^2 2\theta}\ll 1\,.
\end{equation}

From Eq.~(\ref{two32}) we obtain the following probability for the
$\nu_\alpha \to \nu_{\alpha'}$ transition in the adiabatic
approximation:
\begin{equationarray}
P(\nu_\alpha \to \nu_{\alpha'}) &&= \sum_i |U^m_{{\alpha'}i}(x)|^2
 |U^m_{{\alpha}i}(x_0)|^2 +
\label{two88}\\
&&+  2 Re \sum_{i<k} U^m_{{\alpha'}i}(x) {U^m_{{\alpha'}k}}^*
e^{- i \int_{x_0}^x{(E_i^m - E^m_k)\, dx}} {U^m_{{\alpha}i}}^*(x_0)
U^m_{{\alpha}k}(x_0)\,.
\nonumber
\end{equationarray}
For solar neutrinos the second term in the r.h.s. of this expression
disappears due to averaging over the energy and the region in 
which neutrinos are produced. Hence
for the averaged transition probability we have 
\begin{equation}
\overline{P}(\nu_\alpha \to \nu_{\alpha'}) = \sum_i 
|U^m_{{\alpha'}i}(x)|^2 |U^m_{{\alpha}i}(x_0)|^2
\label{two77}
\end{equation}
Thus, in the adiabatic approximation, the averaged transition
probability is determined by the elements of the mixing 
matrix in matter at the initial and final points.
For the case of two neutrino flavors we have the 
following simple expression for the $\nu_e$ survival probability
\begin{equationarray}
\overline{P}(\nu_e \to \nu_e) &&= \cos^2 \theta^m(x) \cos^2 \theta^m(x_0) 
+ \sin^2 \theta^m(x) \sin^2 \theta^m(x_0)
\nonumber \\
&&= \frac {1}{2}\left(1 + \cos 2\theta^m(x) \cos 2\theta^m(x_0)\right) 
\label{two66}
\end{equationarray}
From Eq.~(\ref{two66}) it is easy to see that if the
 neutrino passes the point $x = x_R$ 
where the resonance condition is satisfied, a large effect of 
disappearance of $\nu_e$ will be observed. 
In fact, the condition (\ref{two99}) is fulfilled if $\cos 2\theta > 0$ 
(neutrino masses are labelled in such a way that $\Delta m^2 > 0$).
At the production point $x_0$ the density is larger than at point $x_R$
and $A(x_0) > \Delta m^2 \cos 2\theta$.
From (\ref{two12}) it follows than $\cos 2\theta(x_0)< 0$.
Thus, if the resonance condition is fulfilled, we see from
Eq.~(\ref{two66}) that $P(\nu_e \to \nu_e) < \frac {1} {2}$.
If the condition
\begin{equation}
A(x_0) \gg \Delta m^2
\end{equation}
is satisfied for neutrinos produced in the center of the Sun,then
$\cos 2\theta^m(x_0) \simeq - 1 $ and, for neutrinos passing through
the Sun, the survival probability is equal to:
\begin{equation}
\overline{P}(\nu_e \to \nu_e) \simeq \frac {1} {2} (1 - \cos 2\theta)
\end{equation}
It is obvious from this expression that the $\nu_e$ survival probability at
small $\theta$ is close to zero:
all $\nu_e$'s are transformed into $\nu_\mu$'s.

Let us consider evolution of neutrino states in such a case.
From Eq.~(\ref{two12}) it follows that, at the production point, 
$\theta^m(x_0) \simeq  {\pi}/{2}$.
From (\ref{two13}) we have then
\begin{equation}
|\nu_e\rangle \simeq |\nu_{2m}\rangle\, ;
\qquad
|\nu_\mu\rangle = -  |\nu_{1m}\rangle 
\qquad (x=x_0)
\end{equation}
Thus at the production point flavor states are states with 
definite energy. In the adiabatic approximation there are no 
transitions between energy levels. In the final point 
$\rho_e$ = 0 and at small $\theta$ we have
\begin{equation}
|\nu_2\rangle \simeq |\nu_{\mu}\rangle ,
\qquad
|\nu_1\rangle \simeq  |\nu_{e}\rangle 
\qquad (x=x_0)
\end{equation}
Thus, all $\nu_e$'s transfer to $\nu_\mu$'s.
The resonance condition (\ref{two99}) was written in units 
$\hbar = c = 1$. We can rewrite it in the following form
\[\Delta m^2 \cos 2\theta \simeq 0.7 \cdot 10^{-7} E \rho {\mathrm{eV}}^2\]
where $\rho$ is the density of matter in g$\cdot$~cm$^{-3}$ and
$E$ is the neutrino energy in MeV. In the central region of the Sun 
$\rho \simeq 10^2\mathrm{ g} \cdot {\mathrm{cm}}^{-3}$ and the energy
of the solar neutrinos is $ \simeq 1 MeV$. 
Thus the resonance condition is satisfied at $\Delta m^2 \simeq
10^{-5}$~eV$^2$.

The expression (\ref{two77}) gives the averaged survival probability
in the adiabatic approximation. In the general case we have
\begin{equation}
\overline{P}(\nu_\alpha \to \nu_{\alpha'}) = 
\sum |U^m_{{\alpha'}i}(x)|^2 P_{ik} |U^m_{{\alpha}k}(x_0)|^2
\label{two55}
\end{equation}
where $P_{ik}$ is the probabilty of transition from the state with
energy $E^m_k$ to the state with energy $E^m_i$.
Let us consider the simplest case of the transition between two types of
neutrinos. From the conservation of the total probability we have
\begin{equation}
P_{11} = 1- P_{21}\, , \qquad
P_{22} = 1 - P_{12}\, ,\qquad
P_{12} = P_{21}
\label{two44}
\end{equation}
Thus in the case of two neutrinos all transition probabilities 
$P_{ik}$ are expressed through $P_{12}$.
With the help of (\ref{two35}), (\ref{two55}) and (\ref{two44}), for the 
$\nu_e$ survival probability we have:
\begin{equation}
\overline{P}(\nu_e \to \nu_e) =
\frac {1} {2} + \left(\frac {1}{2} - P_{12}\right)
\cos 2\theta^m(x) \cos 2\theta^m (x_0)
\end{equation}

In the literature there exist different approximate expressions for 
the transition probability $P_{12}$. In the Landau--Zenner approximation,
based on the assumption that the transition occurs mainly in the resonance 
region,
\begin{equation}
P_{12} = e^{- \frac {\pi} {2} \gamma_R F}
\end{equation}
where
\begin{equation}
\gamma_R = \frac{\frac{1}{2} \Delta E^m}{|{\partial\theta^m}/{\partial x}|}
= \frac{\Delta m^2 \sin^2 2\theta} {2 p \cos 2\theta
|\frac{\partial}{\partial x} \ln \rho_e(x_R)|}\,.
\end{equation}
In the above equation $F = 1$ for linear density and
$F = 1 -\tan^2\theta$ for exponential density.
The adiabatic approximation is valid if $\gamma_R \gg 1$
[see (\ref{two37})]. In this case $P_{12} \simeq 0$. 

This concludes the considerations on the phenomenological 
theory of neutrino mixing and on the theory of neutrino oscillations in
vacuum and in matter. 
We will start now  the discussion of experimental data.
There are three methods to search for the effects of neutrino masses
and mixing: 

\vspace{0.3cm}
\noindent I. The precise measurement of the high energy part of 
$\beta$--spectrum; 

\vspace{0.3cm}
\noindent II. The search for neutrinoless double $\beta$--decay; 

\vspace{0.3cm}
\noindent III. The investigation of neutrino oscillations . 

We shall discuss now the results which have been obtained 
in some of the most recent experiments.

\section{Search for effects of neutrino mass\\
 in experiments on the measurement\\
  of the $\beta$-spectrum of $^3$H}

We will discuss here briefly the results of searching for effects 
of neutrino masses in experiments on the measurement of the high-energy
part of the $\beta$--spectrum in the decay
\begin{equation}
^3\mathrm{H} \to {^3{\mathrm{He}}} + e^- + \overline{\nu_e}
\label{two76}
\end{equation}
The process (\ref{two76}) is a superallowed $\beta$--decay:
the nuclear matrix element is constant and the $\beta$--spectrum
is determined by the phase--space factor 
and the Coulomb interaction of the final $e^-$ and $^3$He. For 
the $\beta$--spectrum we have
\begin{equation}
\frac {dN} {dT} = C\, p E (Q - T) 
\sqrt{(Q - T)^2 - m^2_\nu} F(E)
\label{two75}
\end{equation}
Here $p$ is electron momentum, $E = m_e + T$ is the
total electron energy, 
$Q = m_{^3\mathrm{H}} -m_{^3\mathrm{He}} - m_e \simeq 18.6$~keV is
the energy release, $C=$ const and $F(E)$ is the Fermi function,
which describes the Coulomb interaction of the final particles.
In the Eq. (\ref{two75}) the term $(Q-T)$ is the neutrino
energy (the recoil energy of $^3$He can be neglected) and the neutrino
mass enters through the neutrino momentum
$p_{\nu} = \sqrt{(Q - T)^2 - m_{\nu}^2}$.
Notice that in the derivation of Eq.~(\ref{two75})
the simplest assumption was done that $\nu_e$ is the particle 
with mass $m_\nu$.

The Kurie function is then determined as follows
\begin{equation}
K(T) = \sqrt{\frac {dN} {dt} \frac {1} {p E F(E)}}
= \sqrt{C} \sqrt{(Q - T) \sqrt{(Q - T)^2 - m_\nu^2}}
\label{two74}
\end{equation}
If $m_\nu$ = 0, the Kurie function is the stright line $K(T)=\sqrt{C}(Q- T)$
and $T_{max} = 0$. If $m_\nu \neq 0$ then $T_{max} = Q- m_\nu$
and at small $m_\nu$ the Kurie function deviates from the stright line 
in the region close to the maximum allowed energy. Thus, if $m_\nu \neq 0$
in the end point part of the spectrum a deficit of observed events must be
measured (with respect to the number of events expected at $m_\nu = 0$). 

In experiments on the search for effects of neutrino mass by $^3$H--method 
no positive indications in favour of $m_\nu \neq 0$  were found.
In these experiments some anomalies were observed. First, practically in 
all experiments the best--fit values of $m_\nu^2$  are negative. This means 
that instead of a deficit of events, some excess is observed. 
Second, in the Troitsk experiment  a peak in electron spectrum is 
observed at the distance of a few eV from the end. The position of the 
peak is changed periodically with time. There are no
doubts that new, more precise experiments are necessary. The results
of two running experiments are presented in Table \ref{table2}.

\begin{table}[t]
\begin{center}
\renewcommand{\arraystretch}{1.45}
\setlength{\tabcolsep}{0.5cm}
\refstepcounter{tables}
\label{table2}
Table \ref{table2}.
Neutrino mass from $^3$H experiments.\\
\vspace {12pt}
\begin{tabular}{|lcc|}
\hline
Experiment & $m^2_\nu$ & $m_\nu$ \\
\hline
Troitsk & $-1.0\pm 3.0\pm 2.0~{\mathrm{eV}}^2$ &
$<$\, $ 2.5~{\mathrm{eV}}$ \\
\hline
Mainz & $-0.1\pm 3.8\pm 1.8~{\mathrm{eV}}^2$ &
$<$\, $ 2.8~{\mathrm{eV}}$ \\
\hline
\end{tabular}
\end{center}
\end{table}

\section{Neutrinoless double $\beta$-decay}

The decay 
\begin{equation}
(A,Z) \to (A,Z+2) + e^- + e^-
\label{pio01}
\end{equation}
is possible only if the total lepton number $L$ is not conserved,
i.e. if neutrinos with definite masses are Majorana particles.
There are many experiments in which neutrinoless double $\beta$-decay 
($(\beta\beta)_{0\nu}$--decay) of $^{76}$Ge, $^{136}$Xe,
 $^{130}$Te, $^{82}$Se, $^{100}$Mo and other even--even nuclei is 
searched for.

Let consider the process (\ref{pio01}) in the framework 
of neutrino mixing. The standard CC Hamiltonian of the weak interaction
has the form
\begin{equation}
H_I = \frac{G_F} {\sqrt{2}} 2\, \overline{e}_L \gamma^\alpha \nu_{eL} 
j_\alpha + h.c.
\end{equation}
Here $j_\alpha$ is the weak hadronic current and 
\begin{equation}
\nu_{eL}= \sum U_{ei} \nu_{iL}
\end{equation}
where $\nu_i$ is the Majorana neutrino field with mass $m_i$.

The ($\beta\beta)_{0{\nu}}$ decay is a process of second order in $G_F$ 
 with an intermediate virtual neutrino.
Neutrino masses and mixing enter into the neutrino propagator\footnote{
We have used the relation $\nu_i^T = - \nu_i C$  that follows from the 
Majorana condition $\nu_i^C =  C \overline{\nu}_i^T = \nu_i$. It is obvious 
that in the case of Dirac neutrino the propagator is equal to zero.}
\begin{equationarray}
{\nu_{eL}}^{\bullet}(x_1){\nu_{eL}^T}^{\bullet}(x_2) &&= \sum_i U_{ei}^2 
{\nu_{iL}}^{\bullet}(x_1){\nu_{iL}^T}^{\bullet}(x_2) 
 = - \sum U^2_{e i} \frac {(1 - \gamma_5)} {2} {\nu_i}^{\bullet}(x_1) 
{{\overline\nu}_i}^{\bullet}(x_2)
\frac {(1 - \gamma_5)} {2} C
\nonumber \\
&&= - \sum U^2_{e i} \frac {(1 - \gamma_5)} {2} \frac {i} {(2\pi)^4} 
\int{\frac
{e^{- i p (x_1 - x_2)}({\rlap{/}p} + m_i)} {p^2 - m^2_i}\,d^4p 
\frac {(1- \gamma_5)}{2} C}
\end{equationarray}%
Taking into account that
\begin{equation}
 \frac {(1 - \gamma_5)}{2}\,({\rlap{/}p}+m_i)\,
\frac{(1 - \gamma_5)}{2} = m_i  \frac {(1-\gamma_5)}{2}\, ,
\label{kon02}
\end{equation}
we come to the conclusion that the matrix element of 
$(\beta\beta)_{0\nu}$--decay is proportional to\footnote{
The term $m_i^2$ in denominator is small with respect to 
characteristic ${p}$ in nuclei ($\simeq$ 10 MeV)
and can be neglected.} 
\begin{equation}
<m> = \sum U_{e i}^2 m_i
\end{equation}
From (\ref{kon02}) it is evident that the proportionality of the matrix 
element of $(\beta\beta)_{0\nu}$--decay to $<m>$ is due to 
the fact that the standard CC interaction is the left--handed one.
If neutrino masses are equal to zero $(\beta\beta)_{0\nu}$--decay is 
forbidden (conservation of helicity).
Notice that, if there is some small admixture of right--handed currents
in the interaction Hamiltonian, the $L - R$ interference gives a contribution 
proportional to the $\rlap{/}{p}$ term in the neutrino propagator. 
Other mechanisms of $(\beta\beta)_{0\nu}$--decay are also possible 
(SUSY with violation of R-parity ect.). 

In the experiments on the search for $(\beta\beta)_{0\nu}$-decay 
very strong bounds on the life--time of this process were
obtained. The results of some of the latest experiments 
are presented in Table \ref{table3}.
From these data upper bounds for $|<m>|$ can be obtained.
The upper bounds depend on the values of the nuclear matrix elements,
the calculation of which is a complicated problem.
From $^{76}$Ge data it follows 
\begin{equation}
|<m>| < (0.5 - 1)\,\, {\mathrm{eV}}
\end{equation}
In the future experiments on the search for $(\beta\beta)_{0\nu}$--decay
(Heidelberg-Moscow, NEMO, CUORE and others) the sensitivity
$|<m>|\, < 0.1$~eV is planned to be achieved.

\begin{table}[t]
\begin{center}
\renewcommand{\arraystretch}{1.45}
\setlength{\tabcolsep}{0.5cm}
\refstepcounter{tables}
\label{table3}
Table \ref{table3}.
Lower bounds of the life--time $T_{1/2}$ of $(\beta\beta)_{0\nu}$--decay\\
\vspace {12pt}
\begin{tabular}{|lcc|}
\hline
Experiment & Element & Lower bound of $T_{1/2}$ \\
\hline
Heidelberg-Moscow & $^{76}\mathrm{Ge}$ & $>\, 1.6 \times 10^{25}$ y \\
\hline
Caltech-PSI-Neuchatel & $^{136}\mathrm{Xe}$ & $>\, 4.4 \times 10^{23}$ y \\
\hline
Milano & $^{130}\mathrm{Te}$ & $>\, 7.7 \times 10^{22}$ y \\
\hline
\end{tabular}
\end{center}
\end{table}

\section{Neutrino oscillation experiments}

We will discuss now the existing experimental data on the search for 
neutrino oscillations. There exist at present  convincing evidences 
in favour of neutrino oscillations, which were obtained in atmospheric 
neutrino experiments and first of all in the Super--Kamiokande experiment. 
Strong indications in favour of neutrino masses and mixing 
were obtained in all solar neutrino experiments.
Finally, some indications in favour of $\nu_\mu \to \nu_e$ transitions
were obtained in the LSND accelerator experiment.
In many reactor and accelerator short baseline experiments
and in the reactor long baseline experiments CHOOZ  no indication
in favour of neutrino oscillations was found. We will start with
the discussion of the results of solar neutrino experiments.

\subsection{Solar neutrinos}

\begin{figure}[p!]
\setlength{\templength}{\arraycolsep} \setlength{\arraycolsep}{0pt}
\newlength{\newtabularwidth}
\setlength{\newtabularwidth}{0.98\textheight}
\begin{minipage}{0.58\newtabularwidth}
\begin{center}
\null \vspace{2cm} \null
\mbox{\epsfig{file=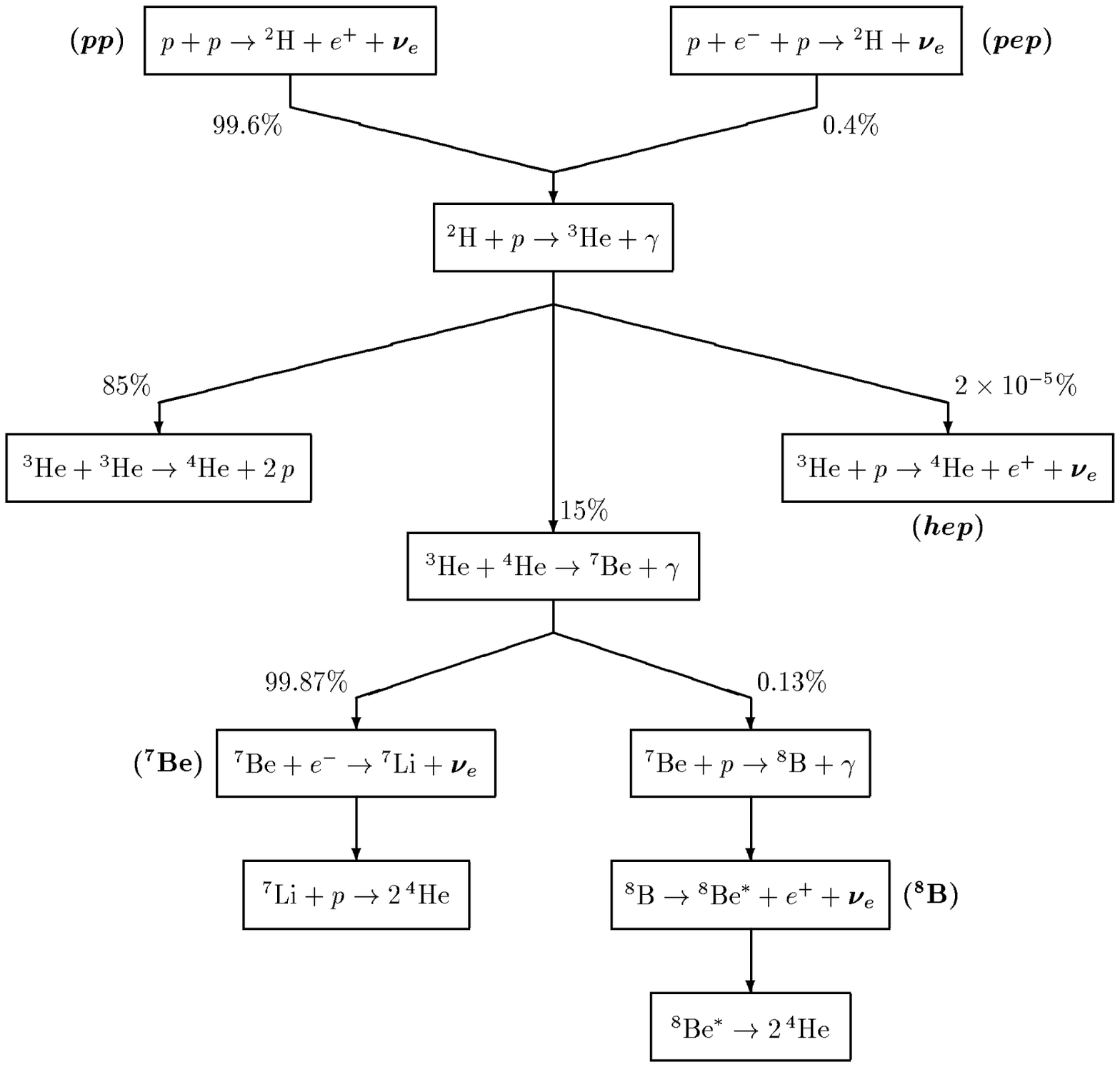,width=0.56\newtabularwidth}}
\end{center}
\end{minipage}
\begin{minipage}{0.48\newtabularwidth}
\refstepcounter{figures}
\label{pp_cycle}
\begin{center}
\null \vspace{1cm} \null
Figure \ref{pp_cycle}.
The $pp$ cycle (the figure is taken from ref.\cite{BGG}).
\end{center}
\end{minipage}
\setlength{\arraycolsep}{\templength}
\end{figure}

The energy of the Sun is generated in the reactions of the thermonuclear 
pp and CNO cycles. The main pp--cycle is illustrated in Fig.\ref{pp_cycle}.
The energy of the sun is produced in the transition
\begin{equation}
4 \, p + 2 \, e^- \to\,{^4}{\mathrm{He}} + 2 \, \nu_e \,,
\label{601}
\end{equation}

If we assume that solar $\nu_e$'s do not transfer into other neutrino
types ($P(\nu_e \to \nu_e) = 1$) we can obtain a relation between the
luminosity of the Sun, ${L}_\odot $ and the flux of solar neutrinos. Let us
consider neutrino with  energy $E$. From (\ref{601}) it follows that 
\begin{equation}
\frac {1} {2} (Q - 2E)
\label{gro}
\end{equation}
is the luminous energy corresponding to the emission of one neutrino.
Here 
\begin{equation}
Q = 4 m_p + 2 m_e - m_{^4{\mathrm{He}}} \simeq 26.7 MeV 
\end{equation}
is the energy release in the transition (\ref{601}). If we multiply 
(\ref{gro} ) by the total flux of solar $\nu_e$'s from different 
reactions and integrate over the neutrino energy $E$
we will obtain the flux of luminous energy from the Sun
\begin{equation}
{\frac{1}{2}\int (Q -2 E) \sum_i I_i(E) dE} =
\frac{L_\odot} {4 \pi R^2}.
\label{kimbo}
\end{equation}  
Here $L_\odot \simeq 3.86 \cdot 10^{33}$ erg/s
is the luminosity of the Sun, $R$ is the Sun--Earth distance and 
$I_i^0(E)$ is the flux of neutrinos from the source $i$ 
($i$ = pp, ...). Notice that in the derivation of the relation 
(\ref{kimbo}) we have assumed that the Sun is in a stationary state.

The luminosity relation (\ref{kimbo}) is solar model independent 
constraint on the solar neutrino fluxes. The flux $I_i(E)$
can be written in the form 
\begin{equation}
I_i(E) =X_i(E)\Phi_i
\end{equation}
where $\Phi_i$ is the total flux and the function $X_i(E)$
describes the form of the spectrum ($\int{X_i(E) dE} = 1$).
The functions $X_i(E)$ are known functions, determined by the
weak interaction. The luminosity relation (\ref{kimbo}) 
can be written in the form 
\begin{equation}
Q \sum_i \left(1- 2\frac{\overline{E}_i} {Q}\right) \Phi_i 
= \frac{L_\odot}{2 \pi R^2}
\end{equation}
where $\overline{E}_i = \int{E X_i(E) dE}$ is the average 
energy of neutrinos from the source $i$. The main sources of solar 
neutrinos are listed in Table \ref{table4}

\begin{table}[t]
\begin{center}
\renewcommand{\arraystretch}{1.45}
\setlength{\tabcolsep}{0.5cm}
\refstepcounter{tables}
\label{table4}
Table \ref{table4}.
Main sources of solar $\nu_e$'s.\\
\vspace {12pt}
\begin{tabular}{|ccc|}
\hline
Reaction  & Maximal energy & Standard Solar Model flux
$({\mathrm{cm}}^{-2}{\mathrm{s}}^{-1})$ \\
\hline
$p\,p \to d\,e^+\,\nu_e$ & $\leqslant 0.42~{\mathrm{MeV}}$ &
$6.0 \times 10^{10}$ \\
\hline
$e^{-}\,{^7}{\mathrm{Be}} \to \nu_e\,{^7}{\mathrm{Li}}$
& $0.86~{\mathrm{MeV}}$ & $4.9 \times 10^9$ \\
\hline
${^8}{\mathrm{B}} \to {^8}{\mathrm{Be}}\,e^+ \,\nu_e$
& $\leqslant 15~{\mathrm{MeV}}$ & $5.0\times 10^6$ \\
\hline
\end{tabular}
\end{center}
\end{table}

As it is seen from the Table, the main source of solar neutrinos is the 
reaction $p + p \to d\,+ e^+\,+ \nu_e$. 
This reaction is the source of low energy neutrinos.
The source of monochromatic medium energy neutrinos is the process 
\begin{equation}
e{^-}+ {{^7}\mathrm{Be}} \to \nu_e + {^7{\mathrm{Li}}}.
\end{equation} 
The reaction 
$^8{\mathrm{B}} \to {^8{\mathrm{Be}}}\, + e^+\,+ \nu_e$ 
is the source of the rare high energy neutrinos.
The results of solar neutrino experiments are presented in 
Table \ref{table5}.

\begin{table}[t]
\begin{center}
\renewcommand{\arraystretch}{1.45}
\setlength{\tabcolsep}{0.5cm}
\refstepcounter{tables}
\label{table5}
Table \ref{table5}.
Results of solar neutrino experiments.\\
\vspace{4pt}
[ $1$ SNU = $10^{-36}$ events/(atoms $\cdot $ sec) ]\\
\vspace {12pt}
\begin{tabular}{|ccc|}
\hline
Experiment & Observed rate & Expected rate \\
\hline
$ \begin{array}{c} {\mathrm{Homestake}}\\ 
\nu_e\,{^{37}\mathrm{Cl}}\to e^-\,{^{37}\mathrm{Ar}}\\
E_{th}=0.81~{\mathrm{MeV}} \end{array}$ &
$2.56 \pm 0.16 \pm 0.16 $ SNU & $7.7 \pm 1.2 $ SNU \\
\hline
$\begin{array}{c} {\mathrm{GALLEX}}\\
\nu_e\,{^{71}\mathrm{Ga}}\to e^-\,{^{71}\mathrm{Ge}}\\
E_{th}=0.23~{\mathrm{MeV}} \end{array}$ &
$ 77.5 \pm 6.2 {}^{+4.3}_{-4.7} $ SNU & $ 129 \pm 8 $ SNU \\
\hline
$\begin{array}{c} {\mathrm{SAGE}}\\
\nu_e\,{^{71}\mathrm{Ga}}\to e^-\,{^{71}\mathrm{Ge}}\\
E_{th}=0.23~{\mathrm{MeV}} \end{array} $ &
$ 66.6 \pm {}^{+6.8~+3.8}_{-7.1~-4.0} $ SNU &
\textemdash~$ \cdot$ ~\textemdash \\
\hline
$\begin{array}{c} {\mathrm{Kamiokande}}\\
\nu e \to \nu e \\
E_{th}=7.0~{\mathrm{MeV}} \end{array} $ &
$(2.80 \pm 0.19 \pm 0.33)~10^6 {\mathrm{cm}}^{-2}{\mathrm{s}}^{-1}$ &
$(5.15 {}^{+1.00}_{-0.72})~10^6 {\mathrm{cm}}^{-2}{\mathrm{s}}^{-1}$ \\
\hline
$\begin{array}{c} {\mathrm{Super-Kamiokande}}\\
\nu e \to \nu e\\ E_{th}=5.5~{\mathrm{MeV}} \end{array} $ &
$(2.44 \pm
0.05{}^{+0.09}_{-0.07})~10^6{\mathrm{cm}}^{-2}{\mathrm{s}}^{-1}$ &
\textemdash~$\cdot$~\textemdash \\
\hline
\end{tabular}
\end{center}
\end{table}

Homestake, GALLEX and SAGE are radiochemical experiments. In the
Kamiokande and the Super--Kamiokande experiments recoil electrons 
(angle and energy)  in the elastic neutrino--electron scattering are 
detected. In these experiments the direction of neutrinos is 
determined and it is confirmed that the detected events are from
solar neutrinos. 

In the Homestake experiment, because of high threshold ($E_{th}=0.81$~MeV)
mainly $^8$B neutrinos are detected: $\simeq 77$\% of events are due 
to $^8$B neutrinos and $\simeq 15$\% of events are due to $^7$Be neutrinos.
In GALLEX and SAGE experiments ($E_{th}=0.23$~MeV)
 neutrinos from all reactions are detected:
$\simeq 54$\% of events are due to $pp$ neutrinos, $\simeq 27$\% of 
events are due to $^7$Be
neutrinos and $\simeq 10$\% of events are due to $^8$B neutrinos.
In the Kamiokande and Super--Kamiokande experiments due to the high 
threshold ($E_{th} =7$~MeV for Kamiokande and $E_{th}=5.5$~MeV
for the Super--Kamiokande) only high energy $^8$B neutrinos are detected.

The results of the solar neutrino experiments are presented in
Table\ref{table5}.  As it is seen from the Table, the detected event rates
in all solar neutrino experiments are significantly
smaller than the predicted one.\footnote{Notice that in the framework of 
neutrino oscillations the possibility of deficit of solar $\nu_e$'s was 
discussed by B.~Pontecorvo in 1968 before the results of the Homestake
experiment were obtained.}
The most natural explanation of the data of solar neutrino experiments can
be obtained in the framework of neutrino mixing.
In fact, if neutrinos are massive and mixed, solar $\nu_e$'s  on 
the way to the earth can transfer into neutrinos of the other types that
are not detected in the radiochemical Homestake, GALLEX and SAGE experiments.
In Kamiokande and Super--Kamiokande experiments all flavor neutrinos 
$\nu_e$, $\nu_\mu$ and $\nu_\tau$ are detected. However, the cross 
section of $\nu_\mu$ ($\nu_\tau$) $- e$ scattering is about six times smaller 
than the cross section of $\nu_e - e$ scattering. 

All existing solar neutrino data 
can be explained if we assume that solar neutrino fluxes are given 
by the Standard Solar Model (SSM) and that there are transitions between 
two neutrino types determined by the two parameters: mass
squared difference 
$\Delta m^2$ and mixing parameter $\sin^2 2\theta$.
We will present the results of such analysis of the data later on. 

Now we will make some remarks about a model independent analysis
of the data. 
First of all from the luminosity relation (\ref{kimbo}) for the 
total flux of solar neutrinos we have the following lower bound
\begin{equation}
\Phi = \sum_i \Phi_i \ge \frac{L_\odot}{2\pi R^2Q}
\end{equation}
Furthermore, for the counting rate in the gallium experiments we have
\begin{equation}
Q_{Ga} = \int_{E_{th}}{\sigma(E) \sum I_i(E) dE}=
\sum_i \overline{\sigma}_i \Phi_i \ge \overline{\sigma}_{pp} \Phi 
= (76 \pm 2)\, {\mathrm{SNU}}
\end{equation}
By comparing this lower bound with the results of the GALLEX and SAGE 
experiments (see Table \ref{table5}) we come to the conclusion that
there is no contradiction 
between experimental data and luminosity constraint if we assume
that there are no transitions of solar neutrinos into other states
($P(\nu_e \to \nu_e) =1$). 

It is possible, however, to show in a model independent
way that the results of {\it different solar neutrino experiments} are not 
compatible if we assume $P(\nu_e \to \nu_e) =1$.
In fact, let us compare the results of the Homestake and the 
Super--Kamiokande experiments. We will consider the total neutrino fluxes
$\Phi_i$ as free parameters.
From the results of Super-Kamiokande experiment we can 
determine the flux of $^8$B neutrinos, $\Phi_{^8B}$
(see Table\ref{table5}).
If we calculate now the contribution of $^8$B neutrinos
into the counting rate of the Homestake experiment we get
\begin{equation}
Q^{^8B}_{Cl} = (2.78 \pm 0.27)\, {\mathrm{SNU}}
\end{equation}
The difference between measured event rate and $Q^{^8B}_{Cl}$
gives the contribution to the Chlorine event rate of $^7$Be and
other neutrinos. We have
\begin{equation}
Q^{^7Be + \dots}_{Cl} = Q^{ex}_{Cl} - Q^{^8B}_{Cl} = (-0.22 \pm 0.35) 
\, {\mathrm{SNU}}
\end{equation}  

All existing solar models predict much larger contribution of $^7$Be
neutrinos to the Chlorine event rate:
\begin{equation}
Q^{^7Be}_{Cl}(SSM) = (1.15 \pm 0.1) SNU
\end{equation} 
The large suppression of the flux of $^7$Be neutrinos
(together with the observation of $^8$B neutrinos) is the problem
for any solar model. The $^8$B nuclei are produced in 
the reaction $p\,+\, {^7{\mathrm{Be}}} \to {^8{\mathrm{B}}}\,+\, \gamma$
and in order to observe neutrinos from $^8$B decay enough $^7$Be nuclei
must exist in the Sun interior. We can come to the same conclusion about 
the suppression of the flux of $^7$Be neutrinos if we compare the results 
of Gallium and Super--Kamiokande experiments.

All existing solar neutrino data can be described if there are 
oscillation between two neutrino flavors, the neutrino fluxes being 
given by the SSM values. If we assume that the oscillation parameters 
$\Delta m^2$ and $\sin^2 2\theta$ are in the region in which matter MSW 
effect can be important, then from the fit of the data two allowed regions 
of the oscillation parameters can be obtained.
For the best fit values it was found
\begin{equationarray}
\Delta m^2\, &= 5 \cdot 10^{-6} \mathrm{eV}^2
\qquad \sin^2 2\theta = 5 \cdot 10^{-3} 
\qquad &\mathrm{(SMA)}
\label{wan4} \\
\Delta m^2 &= 2 \cdot 10^{-5} \mathrm{eV}^2
\qquad \sin^2 2\theta = 0.76       
\qquad &\mathrm{(LMA)}
\label{gro10}
\end{equationarray}
The data can be also described if we assume that the oscillation parameters
are in the region in which matter effects can be neglected
(the case of vacuum oscillations).
For the best fit values it was found in this case 
\begin{equation}
\Delta m^2 = 4.3 \cdot 10^{-10} \mathrm{eV}^2
\qquad \sin^2 2\theta = 0.79          
\qquad \mathrm{(VO)}\, .
\label{gro20}
\end{equation}

In the Super--Kamiokande experiment during 825 days 11240 solar neutrino
events were observed. Such a large statistics allows
the Super--Kamiokande collaboration to measure the energy spectrum of
the recoil electrons and day/night asymmetry. No significant deviation 
from the expected spectrum was observed (may be with 
the exception of the high energy part of the spectrum).
For the day/night asymmetry the following value was obtained 
\begin{equation}
\frac{1}{2}\left( {\frac {N-D} {N+D}}\right) = 0.065 \pm 0.031 \pm 0.013 
\end{equation}
From the analysis of the latest Super--Kamiokande data 
the following best--fit values of the oscillation parameters were found:
\begin{equationarray}
\Delta m^2 &= 5 \cdot 10^{-6} \mathrm{eV}^2
\qquad\quad \sin^2 2\theta = 5 \cdot 10^{-3}
\qquad &\mathrm{(SMA)}
\label{SK1}\\
\Delta m^2 &= 3.2 \cdot 10^{-5}\mathrm{eV}^2
\qquad\, \sin^2 2\theta = 0.8 \qquad &\mathrm{(LMA)}
\label{SK2}\\
\Delta m^2 &= 4.3 \cdot 10^{-10} \mathrm{eV}^2
\qquad \sin^2 2\theta = 0.79 \qquad &{\mathrm{(VO)}}
\label{SK3}
\end{equationarray}
These values are compatible with the ones in Eq. (\ref{wan4}), 
(\ref{gro10}) and (\ref{gro20}), which 
 were found from the analysis of the event rates measured in all
solar neutrino experiments.  

The new solar neutrino experiment SNO started recently in Canada.
The target in this experiment is heavy water (1 kton of $D_2O$)
and Cerenkov light is detected by $\simeq 10^4$ photomultipliers.
Neutrinos will be detected through the observation of the CC reaction
\begin{equation}
\nu_e\,+\, d \to e^-\,+\, p\,+\, p 
\label{rea1}
\end{equation}
as well as of the NC reaction 
\begin{equation}
\nu\, +\, d \to \nu\, +\, n\,+\, p 
\label{rea2}
\end{equation} 
and $\nu - e$ elastic scattering
\begin{equation}
\nu\,+\, e \to \nu\,+\, e 
\end{equation} 
The detection of neutrinos via the CC process (\ref{rea1})
will allow to measure the spectrum of $\nu_e$ on the Earth.
The detection of neutrinos via the NC process (\ref{rea2}) (neutrons will 
be detected) will allow to determine the total flux of flavor 
neutrinos $\nu_e, \nu_\mu, \nu_\tau$. From the comparison of NC and CC 
event rates model independent conclusions on the transition of solar 
$\nu_e$'s into other flavor states can be made.

Next solar neutrino experiment will be BOREXINO. In this experiment
300 tons of liquid scintillator of very high purity will be used.
Solar neutrinos will be detected through the observation of the recoil 
electrons in the process 
\begin{equation}
\nu\,+\, e \to \nu\,+\, e\, .
\end{equation}
The energy threshold in the BOREXINO experiment will be very low, about
250 keV. That will allow
to detect the monoenergetic $^7$Be neutrinos. If vacuum oscillations are 
the origin of the solar neutrino problem, a  seasonal 
variation of the $^7$Be neutrino signal (due to excentricity
of the Earth orbit) will be observed.

\subsection{Atmospheric neutrinos}

Atmospheric neutrinos are produced mainly in the decays of pions
 and muons 
\begin{equation}
\pi \to \mu\,+\, \nu_\mu ,
\qquad
\mu \to e\,+\, \nu_e \,+\,\nu_\mu
\label{rrea1}
\end{equation}
pions being produced in the interaction of cosmic rays in the Earth 
atmosphere. Notice that in the existing detectors neutrino and 
antineutrino events cannot be distinguished.
At small energies, $\le 1$~GeV, the ratio of fluxes of $\nu_\mu$'s
 and $\nu_e$'s from the chain (\ref{rrea1}) is equal to two. 
At the higher energies this ratio is larger than 
two (not all muons decay in the atmosphere) but it can be
predicted with accuracy better than 5\% (the absolute fluxes 
of muon and electron neutrinos are predicted presently with accuracy 
not better than 20 -- 25\%).
This is the reason why the results of the measurements of total fluxes 
of atmospheric neutrinos are presented in the form of a double ratio 
\begin{equation}
R = \frac{\left({N_\mu}/ {N_e}\right)_{\mathrm{data}}}
 {\left({N_\mu}/{N_e}\right)_{\mathrm{MC}}}
\end{equation}
where $({N_\mu}/{N_e})_{\mathrm{data}}$ is the ratio of the total number 
of observed muon and electron events and $({N_\mu}/ {N_e})_{\mathrm{MC}}$ 
is the ratio predicted from Monte Carlo simulations.

We will discuss the results of the Super--Kamiokande experiment. 
A large water Cerenkov detector is used in this experiment. 
The detector consists of two parts: the inner one of 50 kton (22.5 kton 
fiducial volume) is covered with 11146 photomultipliers and the outer
part, 2.75 m thick, is covered with 1885 photomultipliers.
The electrons and muons are detected through the observation of the 
Cerenkov radiation.
The efficiency of particle identification is larger than 98\%. 
The observed events are divided in fully contained events (FC) 
for which Cerenkov light is deposited in the inner detector
and partially contained events (PC) in which the muon track
deposits part of its Cerenkov radiation in the outer detector.
FC events are further divided into sub-GeV events ($E_{vis} \le$ 1.33 
GeV) and multi-GeV events $E_{vis} \geq$ 1.33 GeV).
In the Super-Kamiokande experiment for sub-GeV events and multi-GeV 
events (FC and PC) the following values of the double ratio $R$ were
obtained, respectively (848.3 days):
\begin{equation}
\begin{array}{rl}
R &= 0.680 ^{+0.023} _{-0.022} \pm 0.053
\\ \null \\
R &= 0.678 ^{+0.042} _{-0.039} \pm 0.080
\end{array}
\label{nico}
\end{equation}
These values are in agreement with the values of $R$ obtained in other
water Cerenkov experiments (Kamiokande and IMB)  and in 
the Soudan2 experiment in which the detector is iron calorimeter.
\begin{equationarray}
R&& = 0.65 \pm 0.05 \pm 0.08 \qquad (\mathrm{Kamiokande})
\label{Kam}\\
R &&= 0.54 \pm 0.05 \pm 0.11 \qquad (\mathrm{IMB})
\label{IMB}\\
R &&= 0.61 \pm 0.15 \pm 0.05 \qquad (\mathrm{Soudan2})
\label{Soudan}
\end{equationarray}%
The fact that the double ratio $R$ is significantly less than one is an 
indication in favor of neutrino oscillations.

The important evidence in favour of neutrino oscillations 
was obtained by the Super--Kamiokande collaboration. These data 
were first reported at NEUTRINO98 conference in Japan, in June 1998.
A significant up--down asymmetry of multi--GeV muon events was discovered 
in the Super--Kamiokande experiment. 

For atmospheric neutrinos the distance between production region 
and detector changes from about 20 km for down--going neutrinos 
($\theta = 0$, $\theta$ being the zenith angle) up to about 13,000 km for 
up--going neutrinos ($\theta = \pi$).
In the Super--Kamiokande experiment for the multi--GeV events the zenith
angle $\theta$ can be determined. In fact, charged leptons follow the 
direction of neutrinos (the averaged angle between the charged lepton and 
the neutrino is $15^o - 20^o$). The possible source of the zenith angle 
dependence of neutrino events is the magnetic field of the Earth. However, 
for neutrinos with energies larger than 2 -- 3 GeV, within a few \% no
$\theta$-dependence of neutrino events is expected. 

The Super-Kamiokande collaboration found a significant zenith angle 
dependence of the multi--GeV muon neutrinos. For the integral up--down 
asymmetry of multi--GeV muon neutrinos (FC and PC) 
the following value was obtained 
\begin{equation}
A_\mu = 0.311 \pm 0.043 \pm 0.010
\end{equation}
Here
\begin{equation}
A = \frac {U - D} {U+D}
\label{frac1}
\end{equation}
where ${\it U}$ is the number of up--going neutrinos 
($\cos\theta \le -0.2$) and ${\it D}$ is the number of down--going 
neutrinos (($\cos\theta \geq 0.2$).
No asymmetry of the electron neutrinos was found:
\begin{equation}
A_e = 0.036 \pm 0.067 \pm 0.02
\end{equation}
                
The Super--Kamiokande data can be described if we assume that 
there are $\nu_\mu \to \nu_\tau$ oscillations.
The following best--fit values of the oscillation parameters
were found from the analysis of FC events 
\begin{equation}
\Delta m^2 = 3.05 \cdot 10^{-3} eV^2 ,
\qquad
\sin^2 2\theta = 0.995
\end{equation}
($\chi^2_{min} = 55.4$ at 67 d.o.f.). Let us notice that if we assume 
that there are no oscillations, then in this case $\chi^2 = 177$ at 69 d.o.f.
From the combined analysis of all data it was found
\begin{equation}
\Delta m^2 \simeq (2 - 6) \cdot 10^{-3} eV^2 ,
\qquad
\sin^2 2\theta > 0.84
\end{equation}

If $\nu_\mu \to \nu_s$ oscillations are assumed, at large energies 
matter effects must be important. From the investigation of the high
energy events (PC and upward--going muon events, muons being produced 
by neutrinos in the rock under the detector) the Super--Kamiokande 
collaboration came to the conclusion that
$\nu_\mu \to \nu_s$ oscillations are disfavoured at 95\% C.L.

The range of oscillation parameters which was obtained from the analysis
of the atmospheric neutrino data will be investigated in details in 
long--baseline experiments. The results of the first LBL reactor 
experiment, CHOOZ, were recently published (in this experiment the 
distance between reactors and detector is $\simeq 1$~km). 
No indication in favour ofthe transitions of ${\overline{\nu_e}}$ 
into other states was found in this experiment. For the ratio $R$ of 
the number of measured and expected events it was found 
\begin{equation}
R = 1.01 \pm 2.8\%\, (\mathrm{stat}) \pm 2.7\% \,(\mathrm{syst})
\end{equation}
These data allow to exclude $\Delta m^2 > 7 \cdot 10^{-4} \mathrm{eV}^2$
at $\sin^2 2\theta = 1$ (90\% C.L.).

In LBL Kam-Land experiment  ${\overline{\nu_e}}$'s from reactors 
at the distance of $150 - 200$~km from the detector will be detected.
Neutrino oscillations $\overline{\nu}_e \leftrightarrow \overline{\nu}_x$
with $\Delta m^2 \gsim 10^{-5} \mathrm{eV}^2$ 
and large values of $\sin^2 2\theta$ will be explored. 
The BOREXINO collaboration plans to detect $\overline{\nu}_e$ 
from reactors at the distance of about 800 km from the detector.

The first LBL accelerator experiment K2K is running now. In this
experiment $\nu_\mu$'s with average energy of $1.4$~GeV, produced at 
KEK accelerator, will be detected in the Super--Kamiokande detector 
(at a the distance of about $250 $~km). The disappearance channel 
$\nu_\mu \to \nu_\mu$ and the appearance channel $\nu_\mu \to \nu_e$
will be investigated in detail. This experiment will be sensitive to
$\Delta m^2 \ge 2 \cdot 10^{-3} \mathrm{eV}^2$ at large $\sin^2 2\theta$.

The LBL MINOS experiment between Fermilab and Soudan (the distance is of
about $730$~km) is under the construction. 
In this experiment all the possible channels
of $\nu_\mu$ transitions will be investigated in the atmospheric
neutrino range of $\Delta m^2$.

The LBL CERN-Gran Sasso experiments (the distance is of about $730$~km) 
ICARUS, NOE  and others, are under constraction at CERN and Gran Sasso. The
direct detection of $\tau$'s from $\nu_\mu \to \nu_\tau$
transition will be one of the major goal of these experiments.

\subsection{LSND experiment}

Some indications in favour of $\nu_\mu \leftrightarrow \nu_e$  
oscillations were found in short--baseline LSND accelerator experiment. 
This experiment was done at the Los Alamos linear accelerator
(with protons of $800$~MeV energy). This is a beam--stop experiment: 
most of $\pi^+$'s in the beam, produced by protons, come to a rest in the
target and decay (mainly by $\pi^+ \to \mu^+\, \nu_\mu$); $\mu^+$'s also 
come to a rest in the target and decay by 
$\mu_+ \to e^+\, \nu_e \,\overline{\nu_\mu}$. 
Thus, the beam--stop target is the source of $\nu_\mu,\, \nu_e$ and 
$\overline{\nu}_\mu$ (no $\overline{\nu}_e$ are produced in the decays).

The large scintillator neutrino detector LSND was located at a
distance of about 30 m from the neutrino source. In the detector
$\nu_e$'s were searched for through the observation of the process 
\begin{equation}
\overline{\nu}_e + p \to e^+ + n
\end{equation}
Both $e^+$ and delayed 2.2~MeV $\gamma$'s from the capture 
$n\, p \to d\, \gamma$ were detected.

In the LSND experiment $33.9 \pm 8.0$ events were observed 
in the interval of $e^+$ energies  $30 < E < 60$ MeV.
Assuming that these events are due 
to $\overline{\nu}_\mu \to \overline{\nu}_e$
transitions, for the transition probability it was found
\begin{equation}
P(\overline{\nu}_\mu \to \overline{\nu}_e) = 
(0.31 \pm 0.09 \pm 0.06) \cdot 10^{-3}
\end{equation}
From the analysis of LSND data the allowed region 
in $\sin^2 2\theta - \Delta m^2$ plot was obtained. If the results
of SBL reactor experiments and SBL accelerator experiments on the search
for $\nu_\mu \to \nu_e$ transitions are taken into account for the 
allowed values of the oscillation parameters it was found
\begin{equation}
0.2 \lsim \Delta m^2 \lsim 2 \mathrm{eV}^2 \qquad
2 \cdot 10^{-3} \lsim \sin^2 2\theta \lsim 4 \cdot 10^{-2}
\end{equation}%
The indications in favour of $\nu_\mu \to \nu_e$ 
oscillations obtained in the LSND experiment will be 
checked by BOONE experiment at Fermilab, scheduled for 2001-2002.

\section{Conclusions}

The problem of neutrino masses and mixing is the central problem 
of today's neutrino physics. More than 40 different experiments 
all over the world are dedicated to the investigation of this problem 
and many new experiments are in preparation. The investigation of
the properties of neutrinos is one of the most important direction 
in the search for a new scale in physics. These investigations will 
be very important for the understanding of the origin of tiny neutrino 
masses and neutrino mixing which, according to the existing data,
is very different from CKM quark mixing.

If all existing data will be confirmed by the future experiments 
it would mean that at least four massive neutrinos exist in nature
(in order to to provide three independent neutrino mass squared
differences:
$\Delta m^2_{\mathrm{solar}} \simeq 10^{-5} \mathrm{eV}^2$ (or $10^{-10}$~eV$^2$), 
$\Delta m^2_{\mathrm{atm}} \simeq 10^{-3} \mathrm{eV}^2$ and
$\Delta m^2_{\mathrm{LSND}} \simeq 1 \mathrm{eV}^2$ ).
From the phenomenological analysis of all existing data 
it follows that in the spectrum of masses of four massive neutrinos 
there are two close masses separated by the "large" one, by the 
about 1 eV LSND gap.
Taking into account big--bang nucleosynthesis constraint on the number 
of neutrinos it can be shown  that the dominant transition of the solar 
neutrinos is $\nu_e \to \nu_{\mathrm{sterile}}$ one and the dominant 
transition of the atmospheric neutrinos is $\nu_\mu \to \nu_\tau$.

If the LSND indication in favour of $\nu_\mu \to \nu_e$ 
oscillations will be not confirmed by the future experiments, the
mixing of three massive neutrinos with mass hierarchy is plausible
scenario.

The nature of massive neutrinos (Dirac or Majorana?) can be determined 
from the experiments on the search for neutrinoless double $\beta$-
decay. It can be shown that from the existing neutrino oscillation data 
it follows that effective Majorana mass $<m>$ in the case of three 
massive Majorana neutrinos with mass hierarchy is not larger 
than $10^{-2}$~eV (the present bound is $|<m>| \simeq 0.5$~eV 
and the sensitivity of the next generation of experiments will be 
$|<m>| \simeq 0.1$~eV).

The sensitivity $|<m>| \simeq 10^{-2}$~eV is very important 
problem of experiments on 
the search for neutrinoless double $\beta$-decay.

\vspace{1cm}
I would like to express my deep gratitude to R. Bernabei, W.M. Alberico 
and S. Bilenkaia for their great help in preparing these lecture notes.

\end{document}